\documentclass{rQUF2e}
\usepackage{graphicx}
\usepackage{cite}
\usepackage{natbib}
\usepackage[caption=false]{subfig}
\usepackage{amsmath}

\begin{document}

\title{Maximum Entropy Production Principle for Stock Returns}

\author{Pawe\l{} Fiedor\thanks{Email: Pawel.F.Fiedor@ieee.org}
\affil{Cracow University of Economics\\
Rakowicka 27, 31-510 Krak\'{o}w, Poland}
}

\maketitle

\begin{abstract}
In our previous studies we have investigated the structural complexity of time series describing stock returns on New York's and Warsaw's stock exchanges, by employing two estimators of Shannon's entropy rate based on Lempel--Ziv and Context Tree Weighting algorithms, which were originally used for data compression. Such structural complexity of the time series describing logarithmic stock returns can be used as a measure of the inherent (model--free) predictability of the underlying price formation processes, testing the Efficient--Market Hypothesis in practice. We have also correlated the estimated predictability with the profitability of standard trading algorithms, and found that these do not use the structure inherent in the stock returns to any significant degree. To find a way to use the structural complexity of the stock returns for the purpose of predictions we propose the Maximum Entropy Production Principle as applied to stock returns, and test it on the two mentioned markets, inquiring into whether it is possible to enhance prediction of stock returns based on the structural complexity of these and the mentioned principle.
\end{abstract}


\begin{classcode}G14; G17\end{classcode}

\section{Introduction}

The concepts of entropy and entropy production in non-equilibrium processes formed the basis of a large part of modern physics, and have also created a core of ideological disputes about the evolution of the world in a more broadly defined scientific community. These discussions were started by the ground-breaking contributions made by Clausius, Boltzmann, Gibbs, and Onsager \citep{Clausius:1854,Boltzmann:1866,Gibbs:1902,Onsager:1931}. Nowadays, the properties of entropy in various systems are widely known and studied \citep{Dugdale:1996,Cover:1991}. Much less developed are the concepts around the entropy production behaviour in non-equilibrium processes, even though these are crucial to understanding a wide variety of real-world systems \citep{Prigogine:1947,Keizer:1987}.

Most scientific disciplines strive to find an universal function, the extremum of which would determine the development of the studied system. Such functions have been successfully found only in some cases of relatively simple systems (the Fermat's principle, the principle of least action). Entropy has been seen by many researchers as a quantity describing the progress of non-equilibrium dissipative processes, which may be leading to such universal conclusions. Indeed in 1947 Prigogine proved the minimum entropy production principle. He subsequently applied it to the description of multiple non-equilibrium processes in physics, chemistry and biology \citep{Prigogine:1947,Prigogine:1950,Henin:1974,Prigogine:1987}. There still remains doubt as to whether one can formulate universal entropy principles, which would govern so diverse and dissimilar non-equilibrium processes, however.

Another principle, known as the Maximum Entropy Production Principle (MEPP), is much less known. It may seem as if it is at odds with Prigogine's principle, but this is not the case \citep{Martyushev:2006}. Nonetheless, it has been overshadowed by the more famous twin principle. The MEPP has been proposed and used by several scientists throughout the twentieth century, as they dealt with general theoretical issues of thermodynamics and statistical physics. This principle states that a non-equilibrium system develops so as to maximise its entropy production under present constraints. This principle has found many applications in physics, biology, and other fields \citep{Martyushev:2006,Ozawa:2003,Martyushev:2000,Beretta:2010,Gheorghiu:2001}.

It is reasonably trivial to treat financial markets as non-equilibrium systems \citep{Ingber:1984,Voit:2001}. Additionally, even though they are very complex systems indeed, it is fairly easy to keep track of entropy and entropy production within financial markets, as the price formation processes are describing the time-evolution of these systems. These processes leave a trace in the form of the time series describing prices at various time quanta. These time series are highly non-stationary, thus usually time series containing logarithmic returns are used \citep{Officer:1972,Starica:2005}. This does not change the outlook from the perspective of entropy production however. Having these time series we are able to estimate entropy production within such system (price formation process for a given financial instrument) by estimating the Kolmogorov--Sinai entropy \citep{Sinai:1959,Kolmogorov:1959}, or alternatively all positive Lapunov exponents. Since within the scope of financial markets Lapunov exponents tend not to have a finite value, we measure Kolmogorov--Sinai entropy by estimating Shannon's entropy rate based on discretised time series describing logarithmic returns for studied stocks \citep{Shannon:1948,Cover:1991}.

In our previous studies \citep{Fiedor:2014} we have analysed various markets in this way, and have found that as time quanta at which we measure the prices become shorter, the entropy production becomes smaller also. That is the price changes are more predictable in high frequency analysis (incidentally this is due to volatility clustering being more important in time series describing high frequency returns). We have also studied the relationship between the predictability (as quantified by entropy production) of a given price formation process and the profitability of standard trading algorithms operating on the same processes (stocks). As it happens, these standard trading algorithms do not perform better on the processes (stocks) which are more predictable \citep{Navet:2008,Fiedor:2014}. We postulate that this is because analysts who create these algorithms postulate (after the Efficient--Market Hypothesis) that there is no predictability in the markets, at least in the sense of information theory. Thus the methods do not use the structural complexity inherent in the time series describing prices, due to their design assuming there is no redundancy.

Since the standard methods of guessing future price changes do not use the information contained in entropy analysis of the price formation processes, we try to find out whether we can enhance the prediction simply by using the notion of entropy production itself. In doing so we introduce and partially test in practice the hypothesis of the Maximum Entropy Production Principle applied to financial markets. Of course, this principle states that the system maximises its entropy production under present constraints. There is as yet no theory of financial markets, and thus we do not know the constraints present. But we know that there are many such constraints, and we also know that financial markets are highly complex and very efficient systems, thus obviously a practical test will not find the system maximising entropy at all times. But if we can find that it does tend to follow the maximum entropy path more often than not, then it will provide a partial proof of this hypothesis. A full proof, as is often the case in studying complex systems, will be very hard if not impossible to obtain.

The paper is organised as follows. In Section~2 we present a method used to estimate entropy production in price formation processes, and the setup for testing the Maximum Entropy Production Principle. In Section~3 we apply the presented methodology to real data describing daily and high frequency stock returns from markets in Warsaw and New York. In Section~4 we discuss obtained results. In Section~5 we conclude the study and propose further research.

\section{Methods}

To test the principle of maximum entropy production in financial markets we first need to define the concept of entropy production in the context of financial markets, and in particular stock returns. Here, instead of the thermodynamic approach, we will use the notion of entropy as defined in information theory, since price formation processes are information generating processes. Within this paradigm we can see the growth of Shannon's entropy with regards to the growth of the studied word length, that is Shannon's entropy rate, as an estimator of entropy production. Entropy rate, as defined by Shannon, is an estimator of Kolmogorov--Sinai entropy. The latter is an appropriate tool for accessing entropy for dynamical processes \citep{Sinai:1959,Kolmogorov:1959}.

Thus we need to define the entropy rate and a method of estimating it for practical purposes. The entropy rate, in information theory, is a term derivative to the notion of Shannon's entropy, which measures the amount of uncertainty in a random variable. Shannon's entropy of a random variable $X$ is defined as 
\begin{equation}
	\label{eq:Def_entropy}
H(X) = -\sum_{i} p(x_i) \log_2 p(x_i) 
\end{equation}
summed over all possible outcomes $\{x_i\}$ with their respective probabilities of $p(x_i)$ \citep{Shannon:1948}. 

Claude Shannon has also introduced the entropy rate, which is a generalisation of the notion of entropy for sequences of dependent random variables, and is thus a good measurement of redundancy within studied sequences. For a stationary stochastic process $X = \{X_i\}$ the entropy rate is defined either as
\begin{equation}
	\label{eq:Def_entropy_rate}
	H(X) = \lim_{n \rightarrow \infty} \frac{1}{n} H(X_1, X_2, \dots, X_n)
\end{equation}
or as
\begin{equation}
	\label{eq:Def_entropy_rate_2}
	H(X) = \lim_{n \rightarrow \infty} H(X_n | X_1, X_2, \dots, X_{n-1})
\end{equation}
where \eqref{eq:Def_entropy_rate} holds for all stochastic processes, and \eqref{eq:Def_entropy_rate_2} requires stationarity. Entropy rate can be interpreted as a measure of the uncertainty in a quantity at time $n$ having observed the complete history up to that point. Alternatively it can be seen as the amount of new information created in a unit of time in the studied process \citep{Cover:1991}. Entropy rate can thus be viewed as the maximum rate of information creation which can be processed as price changes for studied financial instruments \citep{Fiedor:2014}. This allows us to use Shannon's entropy rate as an estimator of entropy production within the studied price formation processes.

Entropy estimation within various systems constitutes a very broad and productive field. It has been especially active in the last two decades due to the advances in neurobiology, and the usefulness of information-theoretic constructs to studying many complex systems. Methods of estimating information-theoretic entropy rate can be divided into two separate groups \citep{Gao:2006}:
\begin{enumerate}
\item Maximum likelihood estimators, which analyse the empirical distribution of all words of a given length within the time series. This approach is characterised by exponentially increasing computational requirements for higher word lengths. Consequently, these methods are not practical for analysing long-term relations, which cannot be ignored in economics and finance.
\item Estimators based on data compression algorithms, particularly Lempel--Ziv \citep{Farah:1995,Kontoyiannis:1998a,Lempel:1977} and Context Tree Weighting \citep{Willems:1995,Kennel:2005} algorithms. Both methods are precise even when operating on limited samples \citep{Louchard:1997,Leonardi:2010}, and are thus better equipped to deal with financial data. In this study we use an estimator based on the Lempel--Ziv algorithm.
\end{enumerate}

The Lempel--Ziv algorithm refers back to Kolmogorov, who defined the complexity of a sequence as the size of the smallest binary program which can produce this sequence \citep{Cover:1991}. This definition is purely abstract, thus intermediate measurements are used for practical applications. One of such measures is the Lempel--Ziv algorithm, which tests the randomness of time series, and can also serve as a data compression algorithm (which is its original use). Lempel--Ziv algorithm measures linear complexity of the sequence of symbols and has been first introduced by Jacob Ziv and Abraham Lempel in 1977 \citep{Lempel:1977}. In practice, the algorithm counts the number of patterns in the studied time series, scanning it from left to right. The distribution of these can be used to find the entropy rate later. For example, the complexity of the below sequence
\[s = 101001010010111110\]
is equal to 8, as when scanning it from left to right we finds eight distinct patterns \citep{Doganaksoy:2006}:
\[1|0|10|01|010|0101|11|110|\]
There have been numerous estimators of Shannon's entropy rate created on the basis of the Lempel--Ziv algorithm. In this study we follow earlier studies \citep{Navet:2008,Fiedor:2014} in using the estimator created by Kontoyiannis in 1998 \citep{Kontoyiannis:1998}. It is widely used \citep{Kennel:2005,Navet:2008} and has good statistical properties \citep{Kontoyiannis:1998}. There is a large choice of slightly different variants to choose from \citep{Gao:2008}, none of which have consistently better characteristics than the used estimator. 

Formally, to calculate Shannon's entropy of a random variable $X$, the probability of each possible outcome $p(x_i)$ must be known. For practical purposes such probabilities are usually not known. Then entropy can be estimated by replacing probabilities with relative frequencies from observed data. Estimating the entropy rate of a stochastic process is more complex than estimating entropy, as random variables in stochastic processes are usually interdependent. The mentioned Lempel--Ziv estimator of Shannon's entropy rate is defined as:
\begin{equation}
	\label{eq:LZ_complexity}
	\hat{H} = \frac{n \log_2 n}{\sum_i \Lambda_i}
\end{equation}
where $n$ is the length of the studied time series, and $\Lambda_i$ denotes the length of the shortest substring starting from time $i$ which has not yet been observed prior to time $i$ (between times $1$ and $i-1$). It has been shown that for stationary ergodic processes $\hat{H}(X)$ converges to the entropy rate $H(X)$ with probability of $1$ as $n$ approaches infinity \citep{Kontoyiannis:1998}.

Context Tree Weighting (CTW), mentioned above, is also a data compression algorithm \citep{Willems:1995,Willems:1996,Willems:1998}, which can be interpreted as a Bayesian procedure for estimating the probability of a string generated by a binary tree process \citep{Gao:2008}. This probability can then be used to estimate Shannon's entropy rate. In this paper we use Lempel--Ziv estimator, as we have shown that these two methods give nearly identical results \citep{Fiedor:2014}.

Having defined the methods for entropy rate estimation, we need to comment on the financial data that is to be fed into these algorithms. It is obvious that prices are the most important indicators of the conditions within financial markets. Other measures, for example volatility or realised volatility, may also be used, but are generally seen as less important. Since the time series describing prices themselves are problematic (mostly due to issues with stationarity), we analyse time series containing logarithmic returns. Let us denote the most recent price of the studied financial instrument occurring on time $t$ during the studied period by $p(t)$. Additionally, $\tau$ is the time horizon or price sampling frequency. Then for each stock the logarithmic returns are sampled
\begin{equation}
\rho_{t} = \ln(p(t)) - \ln(p(t-\tau))
\end{equation}
Such time series can describe logarithmic returns of the studied financial instruments at any time horizon, either daily (end of day prices) or intraday. For the purposes of the information-theoretic analysis we need to have discrete time series. Thus we discretise these time series by binning the values into $\Omega$ distinct states. The discrete  logarithmic returns take values from an alphabet with cardinality $\Omega$:
\begin{equation}
r_{t} \in \{1,\dots,\Omega\}
\end{equation}
In our analysis we choose $\Omega=4$, though in our previous studies we have shown that other values give reasonably similar results \citep{Fiedor:2014}. The four states represent four quartiles, and each state is assigned the same number of data points. This means that the model has no unnecessary parameters, which could affect the results and conclusions reached while using the data. Comparable experimental setups have been used in similar studies \citep{Navet:2008,Fiedor:2014} and proved to be very efficient \citep{Steuer:2001}. Here we also note that one could try to use the permutation entropy instead of this approach, which can been seen as a better option for many analyses. Our studies indicate that permutation entropy performs worse for the analysis of financial markets however, and additionally we do not find the assumption that market participants can distinguish between four different levels of market returns too stringent.

Finally, we have to formally introduce the principle of maximum entropy production for financial markets, as well as the setup of the test for the principle of maximum entropy production in practice. In this study we understand entropy production in the sense of Shannon's entropy rate, as defined above. Then if we study particular time series describing logarithmic returns in a window of length $\mu$ ($\{r_{t-\mu},\dots,r_{t}\}$) we can, under the principle of maximum entropy production, say that the next price ($r_{t+1}$) will be assigned the state which maximises Shannon's entropy rate, that is $H(\{r_{t-\mu},\dots,r_{t+1}\})$. This approach does not predict the price exactly, but to the accuracy as specified in the discretisation step, that is differentiating among $\Omega$ different levels. In other words $\alpha,\beta\in\{1,\dots,\Omega\}$. Formally the principle of maximum entropy production for stock returns can be described as:
\begin{equation}
\Psi=P(r_{t+1}=\alpha \; | \; \forall_{\beta \neq \alpha} \; H(\{r_{t-\mu},\dots,r_{t},\alpha\})>H(\{r_{t-\mu},\dots,r_{t},\beta\}))=1
\end{equation}
Given no constraints this approach should give $100\%$ accuracy. But we know that financial markets are highly complex adaptive systems with many (unknown) constraints, and we have shown that they are nearly efficient \citep{Fiedor:2014}, thus in practice we can only see whether the future (out of sample, predicted) state which maximises entropy production is significantly overrepresented in the historical data for various financial markets. Thus the principle of maximum entropy production for stock returns in practice could be described as:
\begin{equation}
\Psi>>\frac{1}{\Omega}
\label{eq:hyp}
\end{equation}
In other words assuming discretisation into four states we want to study financial data with various values of parameter $\mu$ and find whether the percentage of times we can guess the next log return by maximising entropy rate given the history of length $\mu$ significantly exceeds $0.25$, which we would get by guessing it randomly. We will attempt to test this in practice by estimating $\Psi$ as the percentage of correct guesses when applying this procedure to historical data, and moving through the time series with moving window of length $\mu$. These will be further averaged over all studied stocks. If we can show that the above inequality holds in practice it would be the first step to proving that the principle of maximum entropy holds for financial markets. This in turn would be of great importance to market practitioners, and particularly analysts of financial markets.

\section{Experimental Results}

First, we introduce the datasets on which we test the principle of maximum entropy production, and the way in which we discretise them. In this study we use datasets from Warsaw's and New York's stock exchanges. This covers both emerging and mature markets, and the database used is large enough for the results not to be accidental.

We start with end of day time series for 360 securities traded on Warsaw Stock Exchange (GPW) (all securities which were listed for over 1000 consecutive days). This data has been downloaded from DM BO\'S database\footnote{http://bossa.pl/notowania/metastock/} and was up to date as of the 5th of July 2013. The data is transformed in the standard way for analysing price movements, so that the data points are the log ratios between consecutive daily closing prices, and those data points are, for the purpose of the Lempel--Ziv estimator, discretised into four distinct states. For high frequency price changes we use another set of data, that is intraday price changes for 707 securities listed on Warsaw Stock Exchange which had over 2500 price changes recorded (all securities with at least 2500 recorded price changes). Unlike in the daily data, in the intraday time series we have ignored data points where there was no price change, as those are not meaningful for data points separated by very small, and not uniform, time intervals. Most of the data points show no price change and calculating entropy in that manner would be pointless, thus in effect we estimate the predictability of the next price change. On the daily data no price change is meaningful however, and hence were not deleted from the data. Analysis with those removed from daily data gives similar results \citep{Fiedor:2014}.

We also study the daily price time series for 98 securities traded on New York Stock Exchange (NYSE100) (the 2 missing stocks were excluded due to missing data). The data has been downloaded from Google Finance database\footnote{http://www.google.com/finance/} and was up to date as of the 11th of November 2013, going 10 years back. The data is transformed in the same way as above. For high frequency price changes we use another set of data, that is intraday (1-minute intervals) price changes for the same 98 securities listed on NYSE 100. The data covers 15 days between the 21st October 2013 and the 8th of November 2013. Therefore the length of time series in both cases are comparable and sufficient for the used algorithms.

Here we note that since in the prediction we have to use two times series, one without the value that we predict ($\{r_{t-\mu},\dots,r_{t}\}$) and one including this value ($\{r_{t-\mu},\dots,r_{t+1}\}$). This raises a problem for the discretisation step. Since the length for these two samples is different the resulting quartiles may also be different. Additionally, if we want to check whether the prediction is correct we may run into an issue where the real values at time $t+1$ is between quartiles, and we will have no good rule to decide on which of the bordering states should this value be assigned. For simplicity, in this study we have discretised the entire time series of length $N$, and not the subseries of length $\mu$ that are under study in each step as we move though the data. This is potentially problematic as we take into account future data. Nonetheless, such quartile discretisation is quite stable, and it also represents the analysts' concentration on what different levels of log returns mean in the long run. Thus we do not find this approach troubling. But to be cautious we need to consider a small correction to the principle of maximum entropy production for stock returns as defined above. If we use a global discretisation step then for some cases we may find that the predicted state for time $t+1$ which maximises entropy production may coincide with the state which is under-represented in the studied subseries. Then this state would be over-represented in the rest of the studied times series, increasing the likelihood of guessing it by random chance over the standard $1/\Omega$. Therefore, to be cautious, we need to compare the results to what we can reasonably get by randomly guessing the future value of logarithmic returns, taking into account the worst case scenario of described over-representation of the guessed state in the studied time series, for a given length of the time series, and a given value of the window $\mu$. Then the principle of maximum entropy production for stock returns is defined as:
\begin{equation}
\Psi>>\frac{N/\Omega}{N-\mu}
\label{eq:hyp2}
\end{equation}
As can be seen, this correction is increasing with increasing ratio $\mu/\Omega$, thus we are able to find out whether this problem affects our analysis by observing how the results change with changing the parameter $\mu$. In this study we use $\mu$ between 20 and 100. There is not much sense in estimating Shannon's entropy rate for time series shorter than 20, as the statistical noise will be significant for such calculations. On the other hand, we are not interested in using very long time series, thus there is no need to analyse values of $\mu$ higher than 100. For the above correction we will use $N$ equal to the average within the studied set of stocks.

\begin{figure*}%
\centering
\subfloat[][]{%
\label{fig:WD4-a}%
\includegraphics[width=0.5\textwidth]{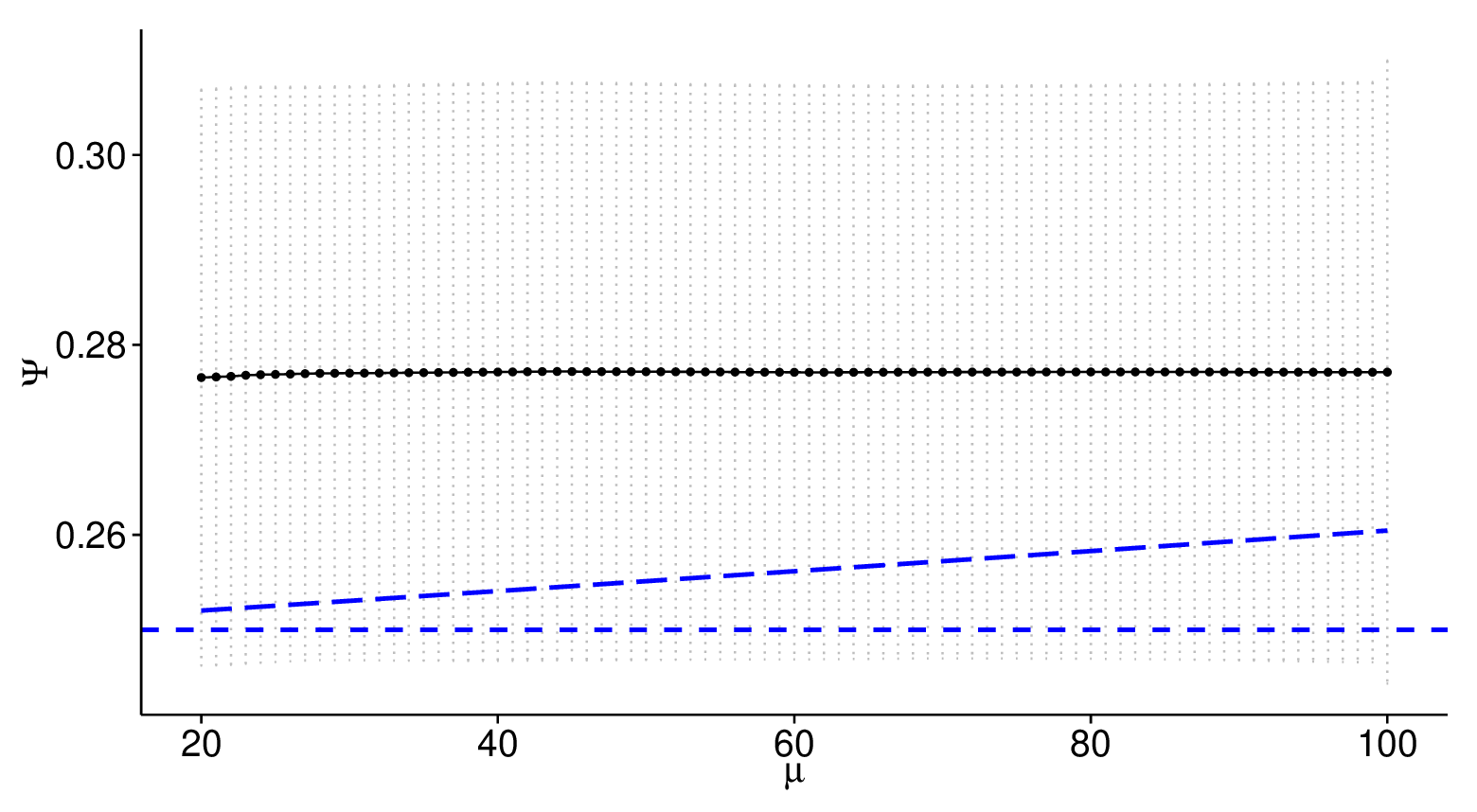}}%
\subfloat[][]{%
\label{fig:WD4-b}%
\includegraphics[width=0.5\textwidth]{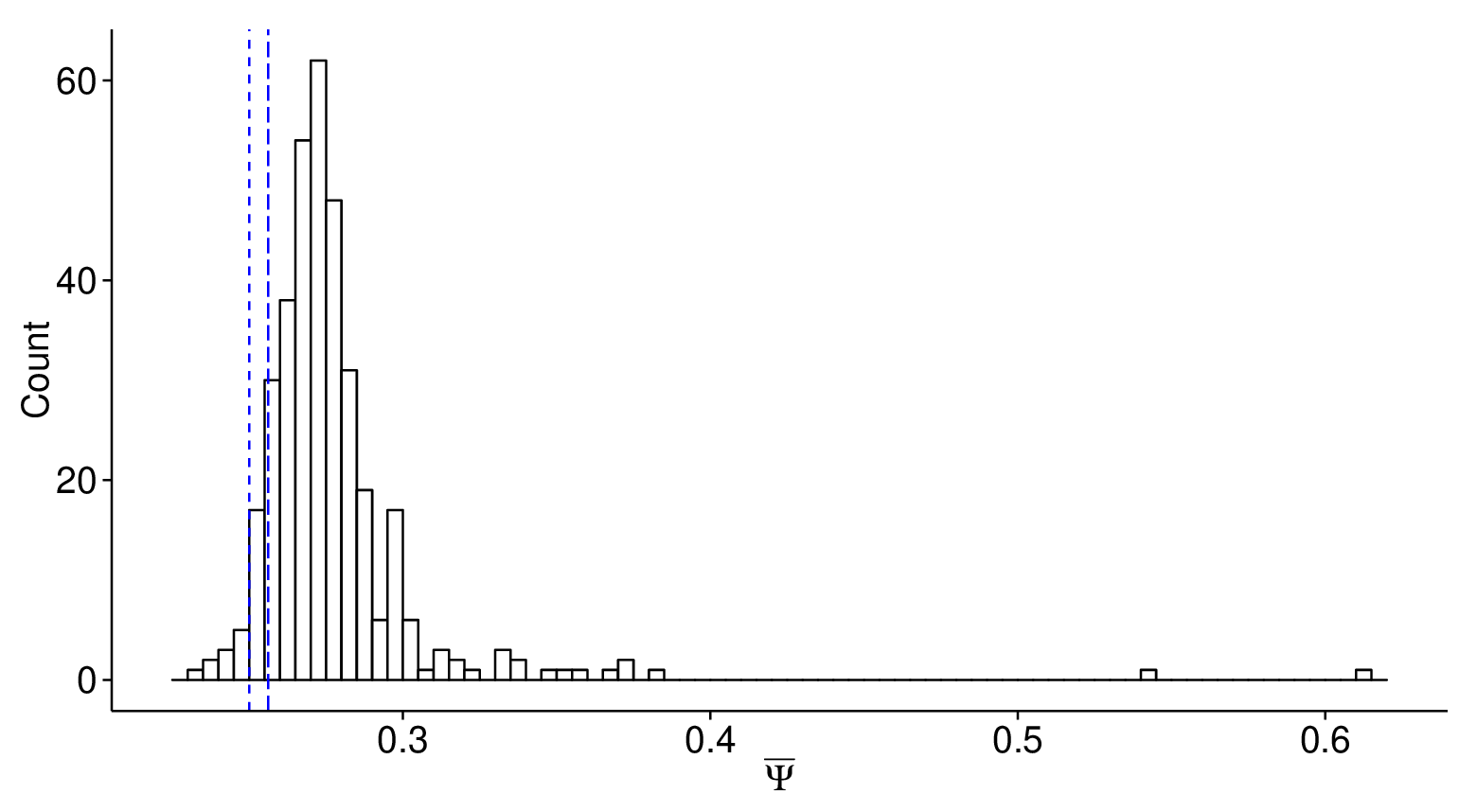}}\\
\subfloat[][]{%
\label{fig:WD4-c}%
\includegraphics[width=0.5\textwidth]{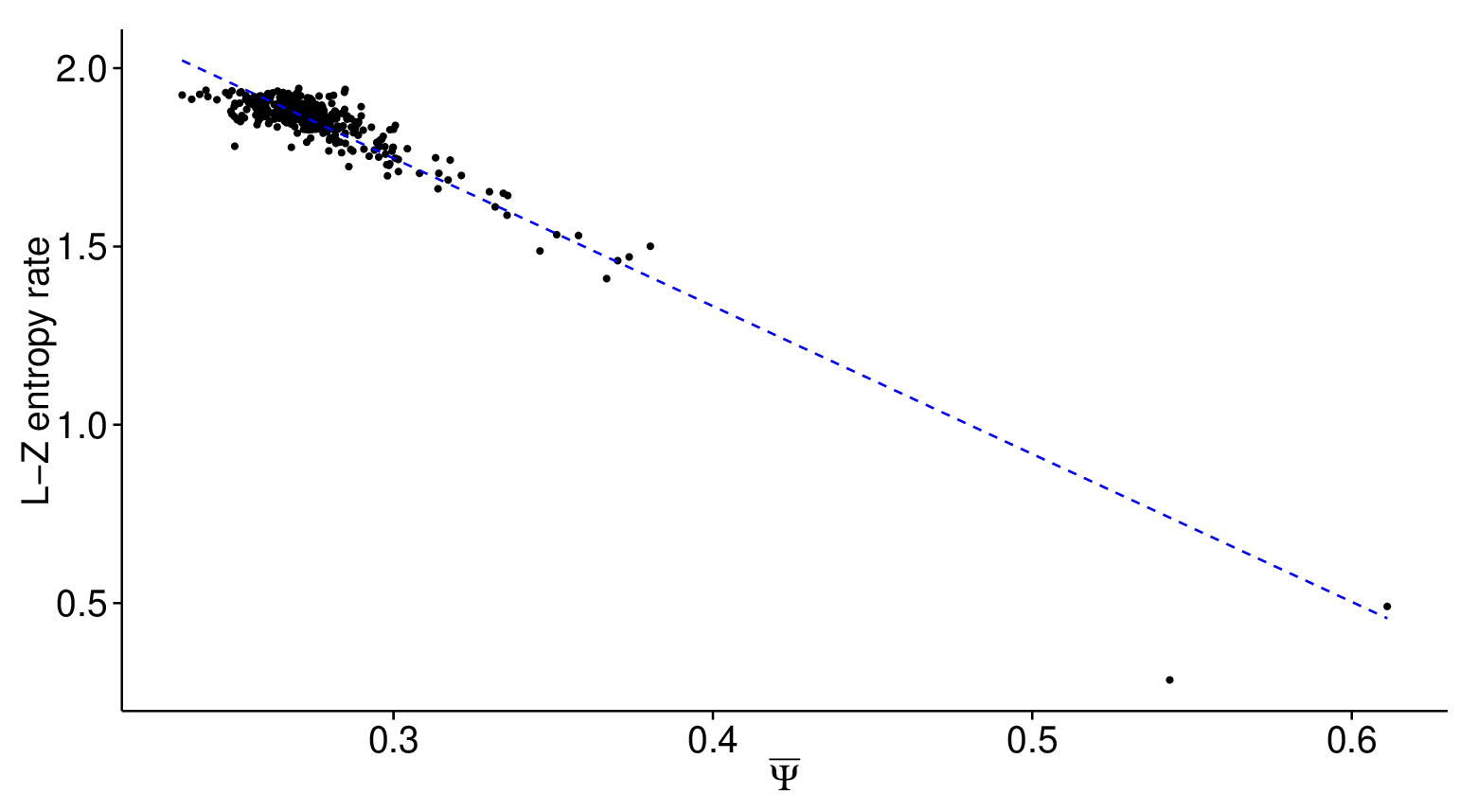}}%
\subfloat[][]{%
\label{fig:WD4-d}%
\includegraphics[width=0.5\textwidth]{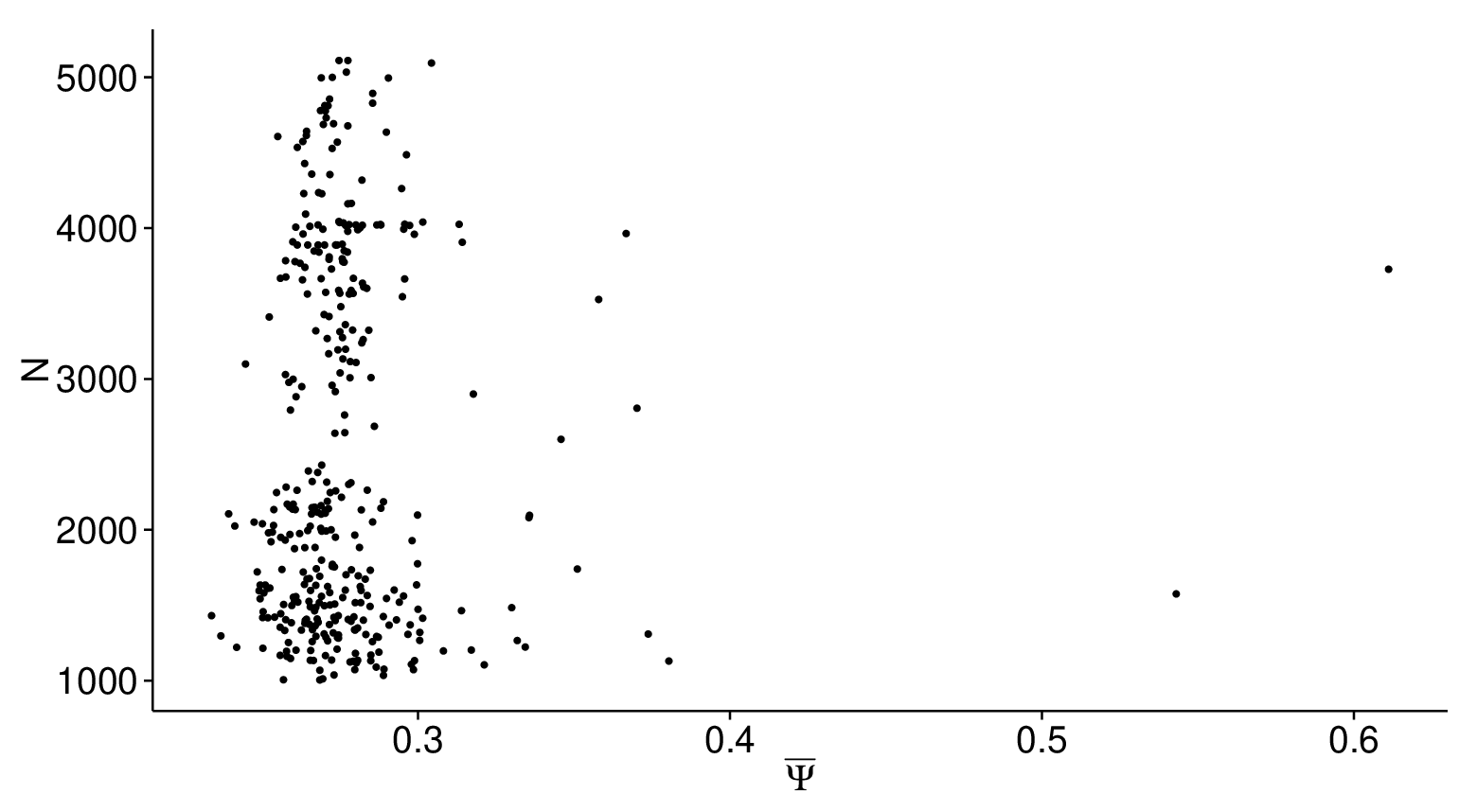}}%
\caption[WD4d]{Maximum Entropy Production Principle (MEPP) testing results for Warsaw Stock Exchange daily logarithmic returns (discretised into four quartiles) database of 360 stocks:
\subref{fig:WD4-a} Estimate of the percentage of times the MEPP correctly predicts the next price move ($\Psi$) averaged over all studied stocks vs the window length parameter ($\mu$), together with error bars of one standard deviation, and two lines denoting values for random guessing the next price move (below unadjusted, above adjusted);
\subref{fig:WD4-b} Histogram of $\Psi$ averaged over all values of $\mu$ together with vertical lines for unadjusted (left) and adjusted (right) values for random guessing;
\subref{fig:WD4-c} Scatterplot of $\Psi$ averaged over all values of $\mu$ vs Lempel--Ziv entropy rate for the whole time series, with linear regression fitted;
\subref{fig:WD4-d} Scatterplot of $\Psi$ averaged over all values of $\mu$ vs the length of the whole time series. Both the average and a great majority of individual stocks are above the values for random chance, supporting our hypothesis. Additionally, the values of $\Psi$ are strongly positively correlated with the predictability of the price formation processes, but the results do not depend on the size of the studied time series.}%
\label{fig:WD4}%
\end{figure*}

In figure~\ref{fig:WD4} we show the results of testing the Maximum Entropy Production Principle for Warsaw Stock Exchange daily logarithmic returns (discretised into four quartiles) database of 360 stocks. In figure~\ref{fig:WD4-a} we show an estimate of the percentage of times the MEPP correctly predicts the next price move ($\Psi$) averaged over all studied stocks with regards to the window length ($\mu$) for which we employ the MEPP. This estimate is performed by moving through the whole studied series with the running window of size $\mu$, and finding the percentage of times we guess the next price movement correctly (when using the MEPP). Within this figure we also show error bars of plus and minus one standard deviation. Further, the two lines denote the theoretic percentages of times we would expect to guess the next price movement by pure chance. The line below (horizontal) is unadjusted, or in other words $1/\Omega$, and the line above is adjusted for our setup as explained above. These results allow us to test the hypothesis as outlined in equations~\eqref{eq:hyp} and \eqref{eq:hyp2}, for all studied values of the window length. These also allow us to see whether the window length affects the results, which can hint as to whether the adjustment mentioned above is necessary. In figure~\ref{fig:WD4-b} we show a histogram of the percentage of times the MEPP correctly predicts the next price move ($\Psi$) averaged over all values of $\mu$, together with vertical lines for unadjusted (on the left) and adjusted (on the right) theoretic percentages of times we would expect to guess the next price movement by chance. This allows us to clearly see how many of the studied stocks fall below the value expected for random guessing, helping to assess the stability of our results. In figure~\ref{fig:WD4-c} we show a scatterplot of the percentage of times the MEPP correctly predicts the next price move ($\Psi$) averaged over all values of $\mu$ with regards to the Lempel--Ziv entropy rate (an estimator of the predictability of the price formation processes) for the whole studied time series. This scatterplot also features a linear regression fitted to the data, to present the correlation between the two variables. This scatterplot allows us to see whether predictions based on the MEPP are working better for more predictable price formation processes, which is not the case for standard methods of prediction used in trading algorithms. In figure~\ref{fig:WD4-d} we show a scatterplot of the percentage of times the MEPP correctly predicts the next price move ($\Psi$) averaged over all values of $\mu$ with regards to the length of the whole studied time series. This scatterplot once again helps us to see whether there is a need for adjusting the testing threshold.

\begin{figure*}%
\centering
\subfloat[][]{%
\label{fig:WI4-a}%
\includegraphics[width=0.5\textwidth]{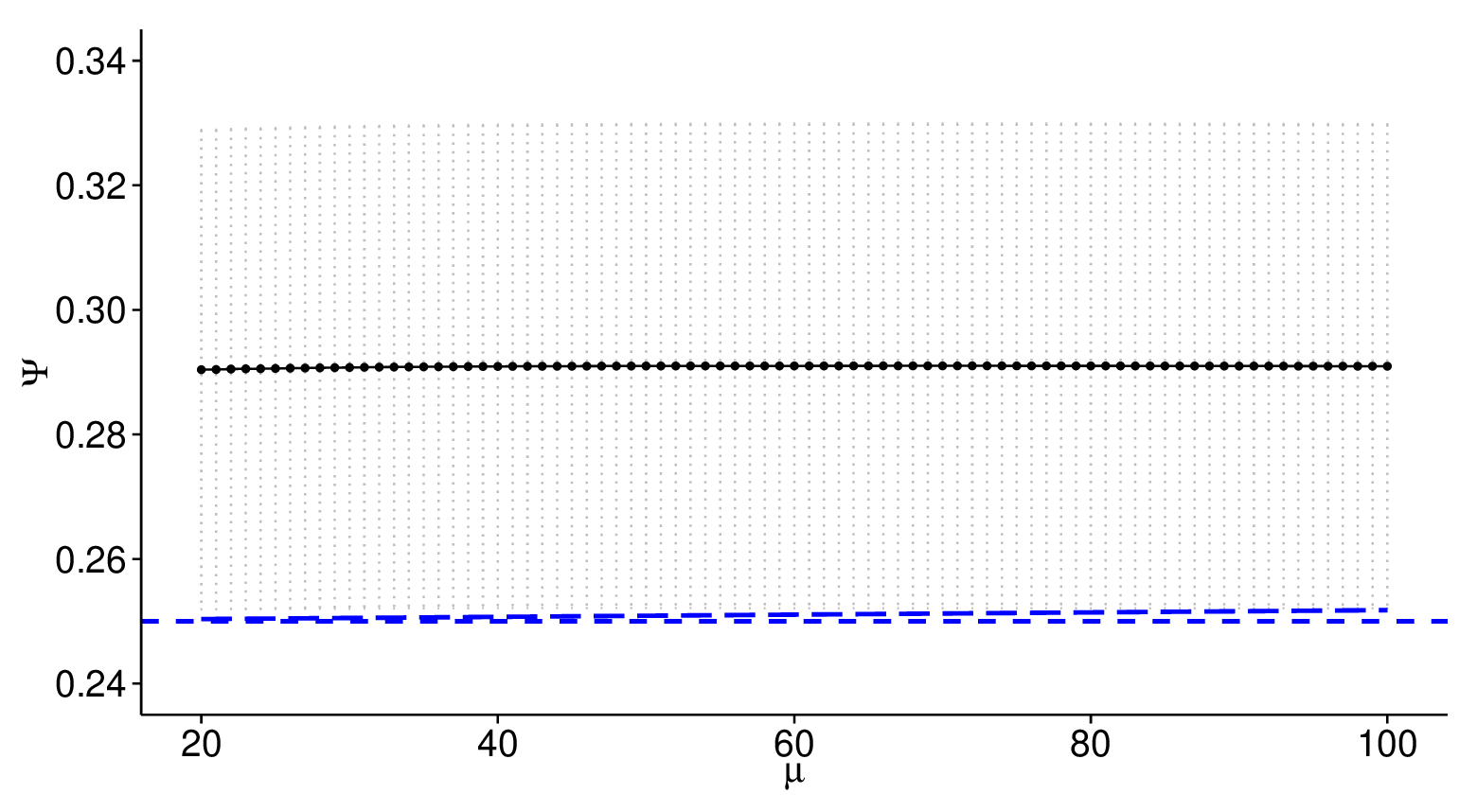}}%
\subfloat[][]{%
\label{fig:WI4-b}%
\includegraphics[width=0.5\textwidth]{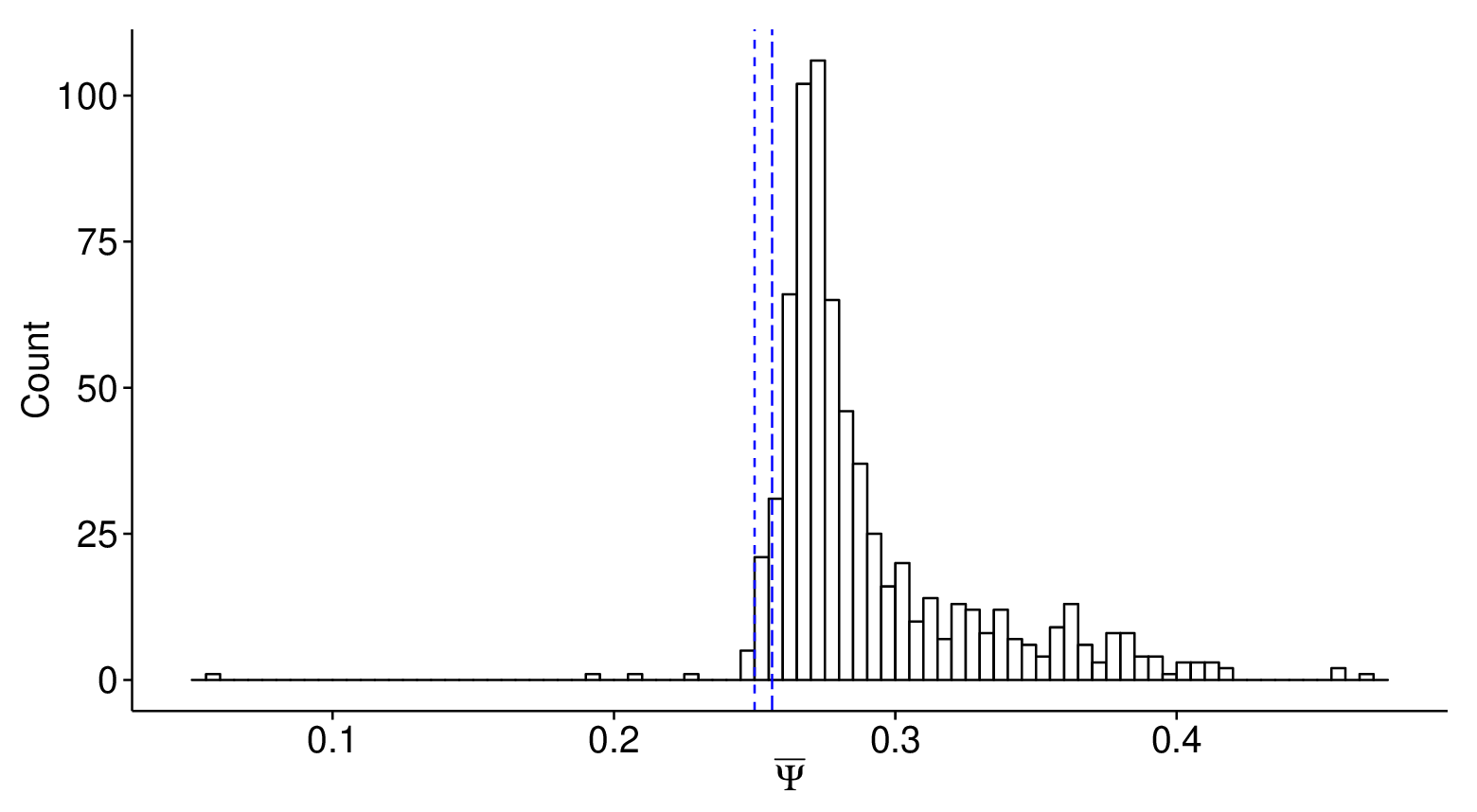}}\\
\subfloat[][]{%
\label{fig:WI4-c}%
\includegraphics[width=0.5\textwidth]{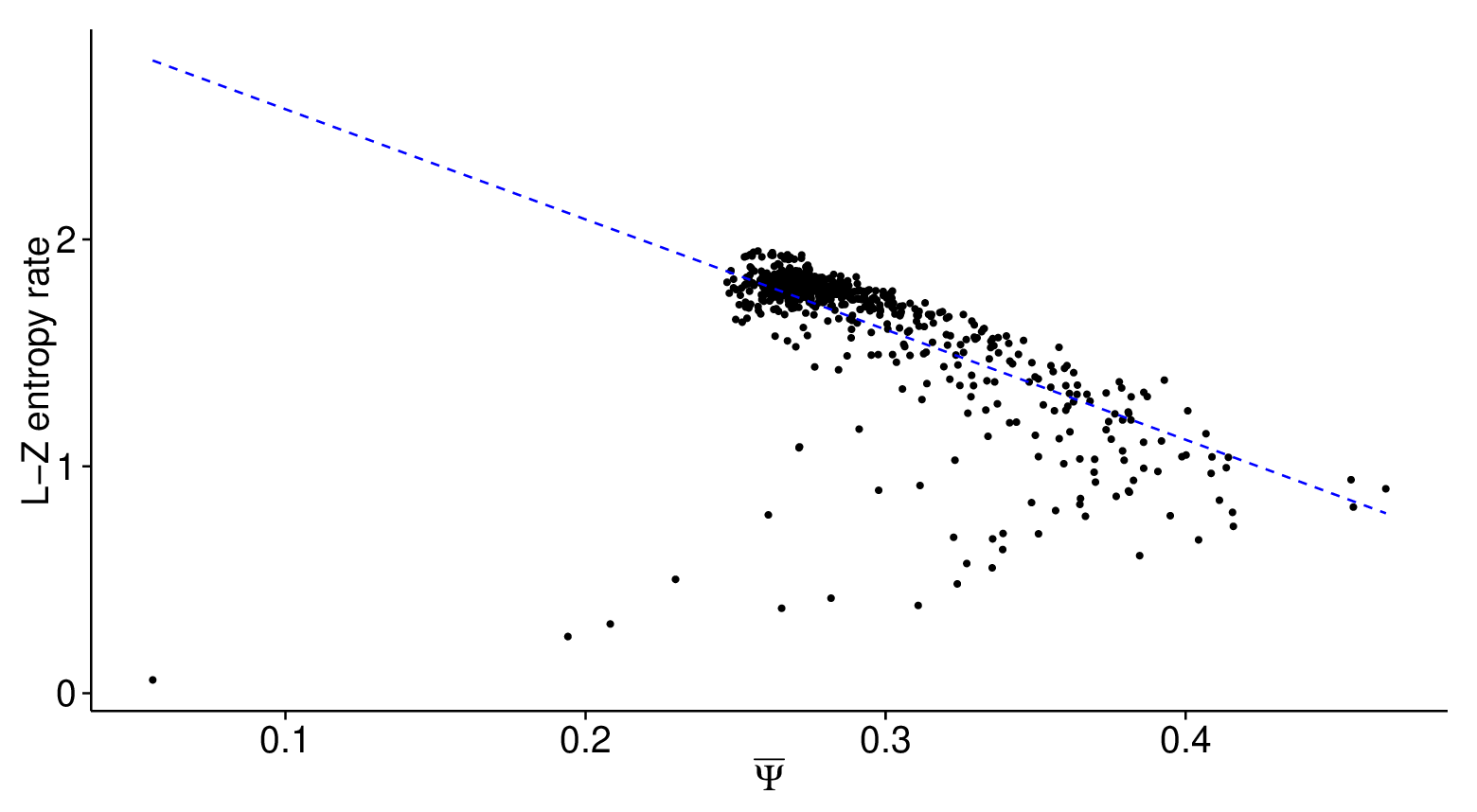}}%
\subfloat[][]{%
\label{fig:WI4-d}%
\includegraphics[width=0.5\textwidth]{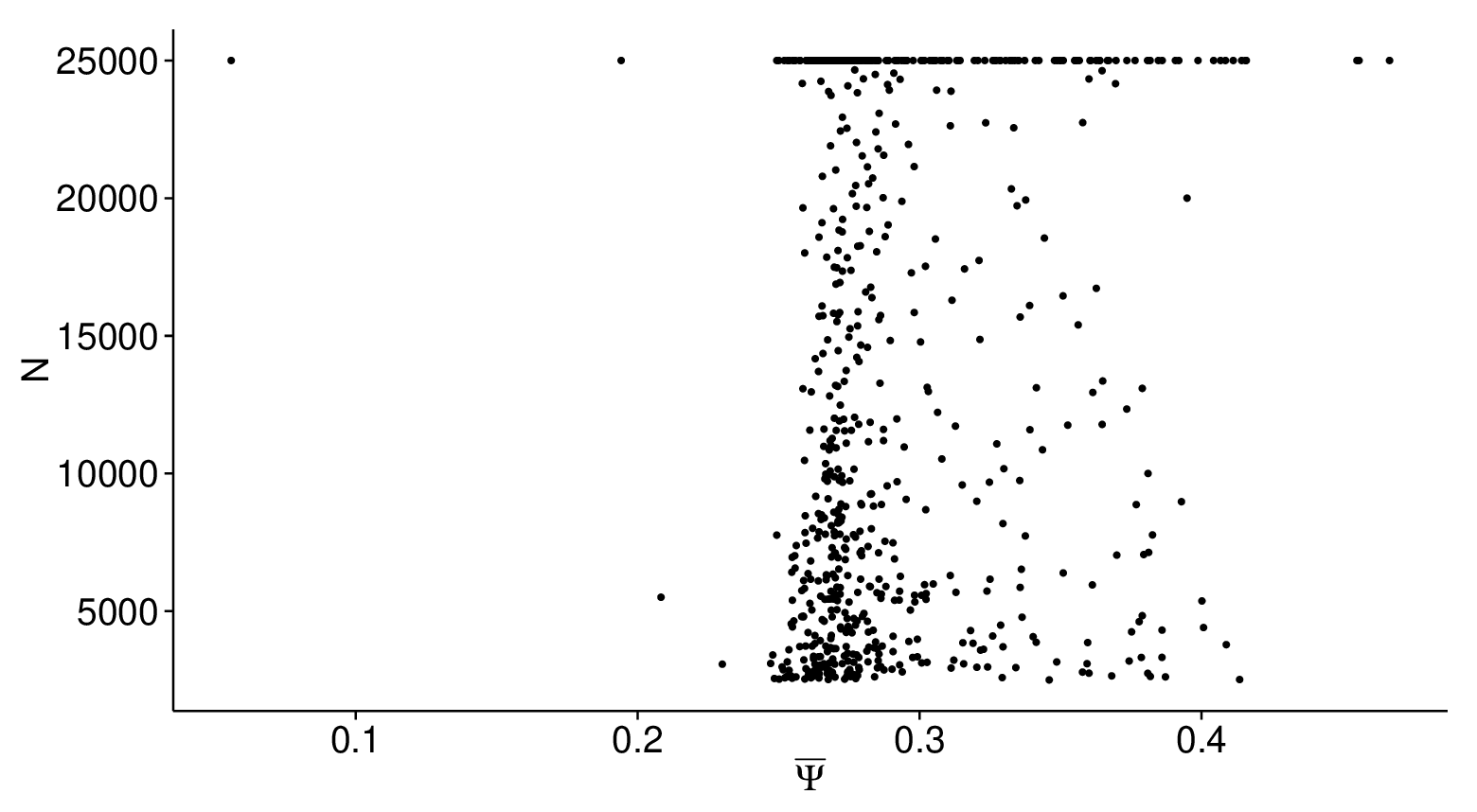}}%
\caption[WI4d]{Maximum Entropy Production Principle (MEPP) testing results for Warsaw Stock Exchange intraday logarithmic returns (discretised into four quartiles) database of 707 stocks:
\subref{fig:WI4-a} Estimate of the percentage of times the MEPP correctly predicts the next price move ($\Psi$) averaged over all studied stocks vs the window length parameter ($\mu$), together with error bars of one standard deviation, and two lines denoting values for random guessing the next price move (below unadjusted, above adjusted);
\subref{fig:WI4-b} Histogram of $\Psi$ averaged over all values of $\mu$ together with vertical lines for unadjusted (left) and adjusted (right) values for random guessing;
\subref{fig:WI4-c} Scatterplot of $\Psi$ averaged over all values of $\mu$ vs Lempel--Ziv entropy rate for the whole time series, with linear regression fitted;
\subref{fig:WI4-d} Scatterplot of $\Psi$ averaged over all values of $\mu$ vs the length of the whole time series. Both the average and a great majority of individual stocks are above the values for random chance, supporting our hypothesis. Additionally, the values of $\Psi$ are strongly positively correlated with the predictability of the price formation processes, but the results do not depend on the size of the studied time series.}%
\label{fig:WI4}%
\end{figure*}

\begin{figure*}%
\centering
\subfloat[][]{%
\label{fig:ND4-a}%
\includegraphics[width=0.5\textwidth]{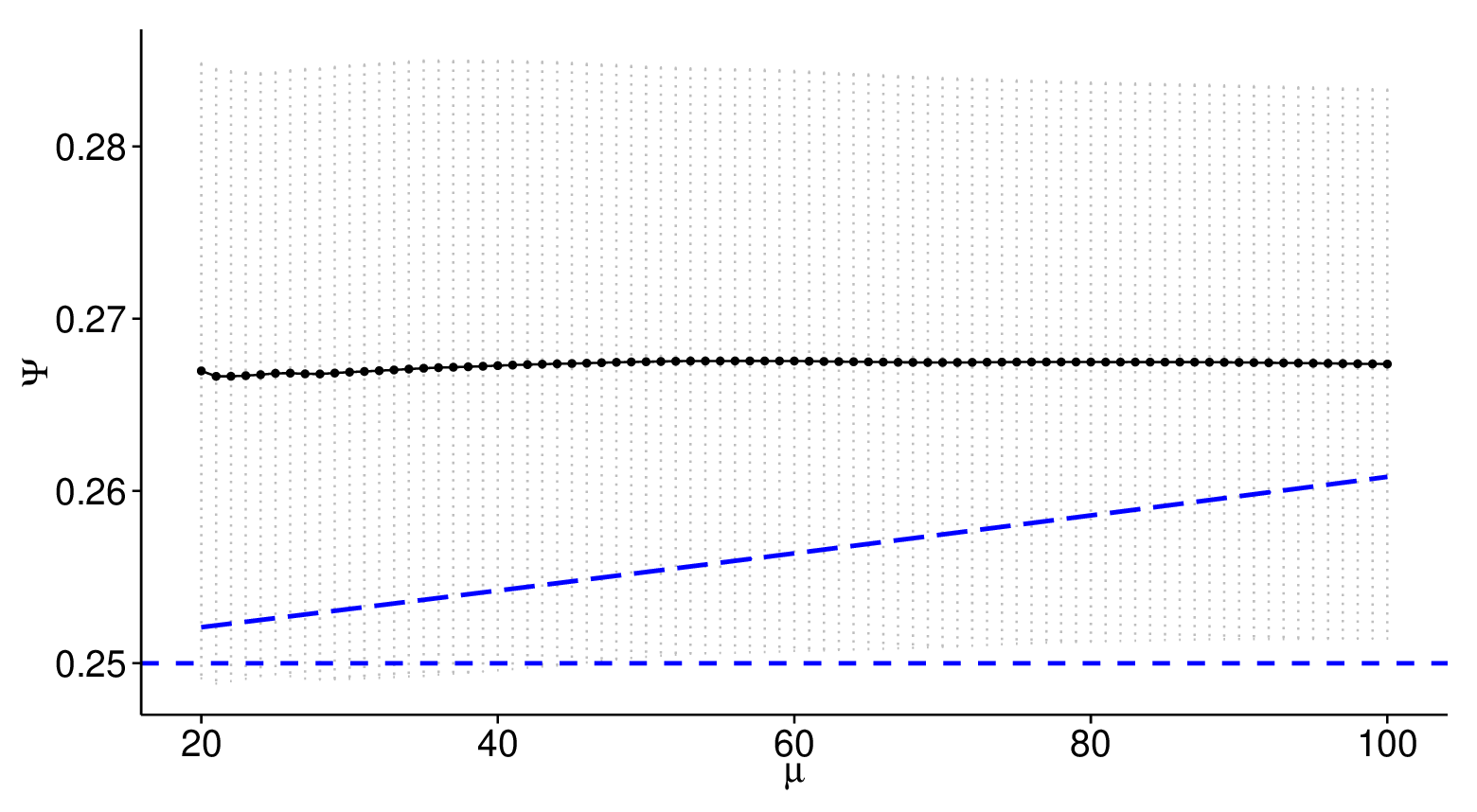}}%
\subfloat[][]{%
\label{fig:ND4-b}%
\includegraphics[width=0.5\textwidth]{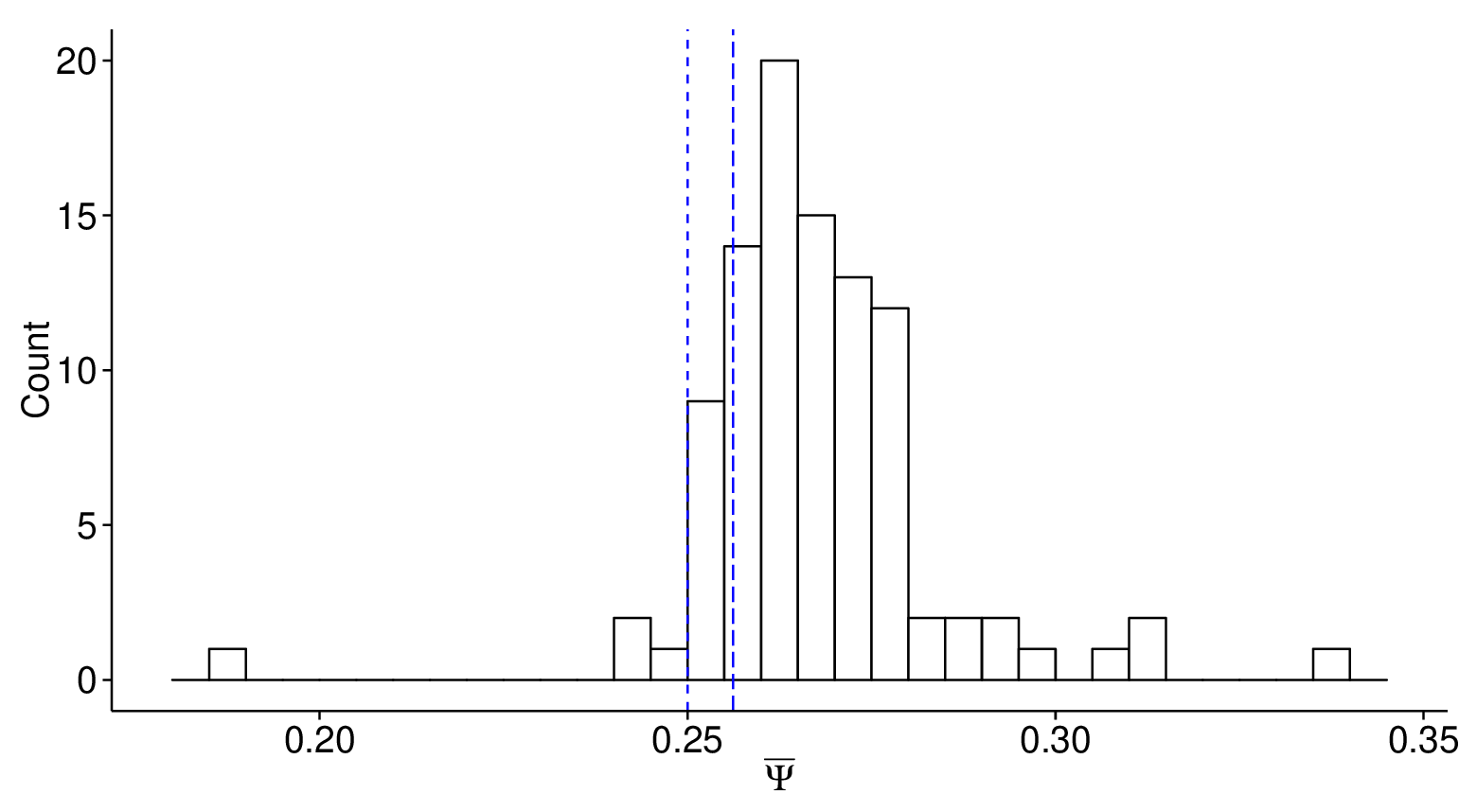}}\\
\subfloat[][]{%
\label{fig:ND4-c}%
\includegraphics[width=0.5\textwidth]{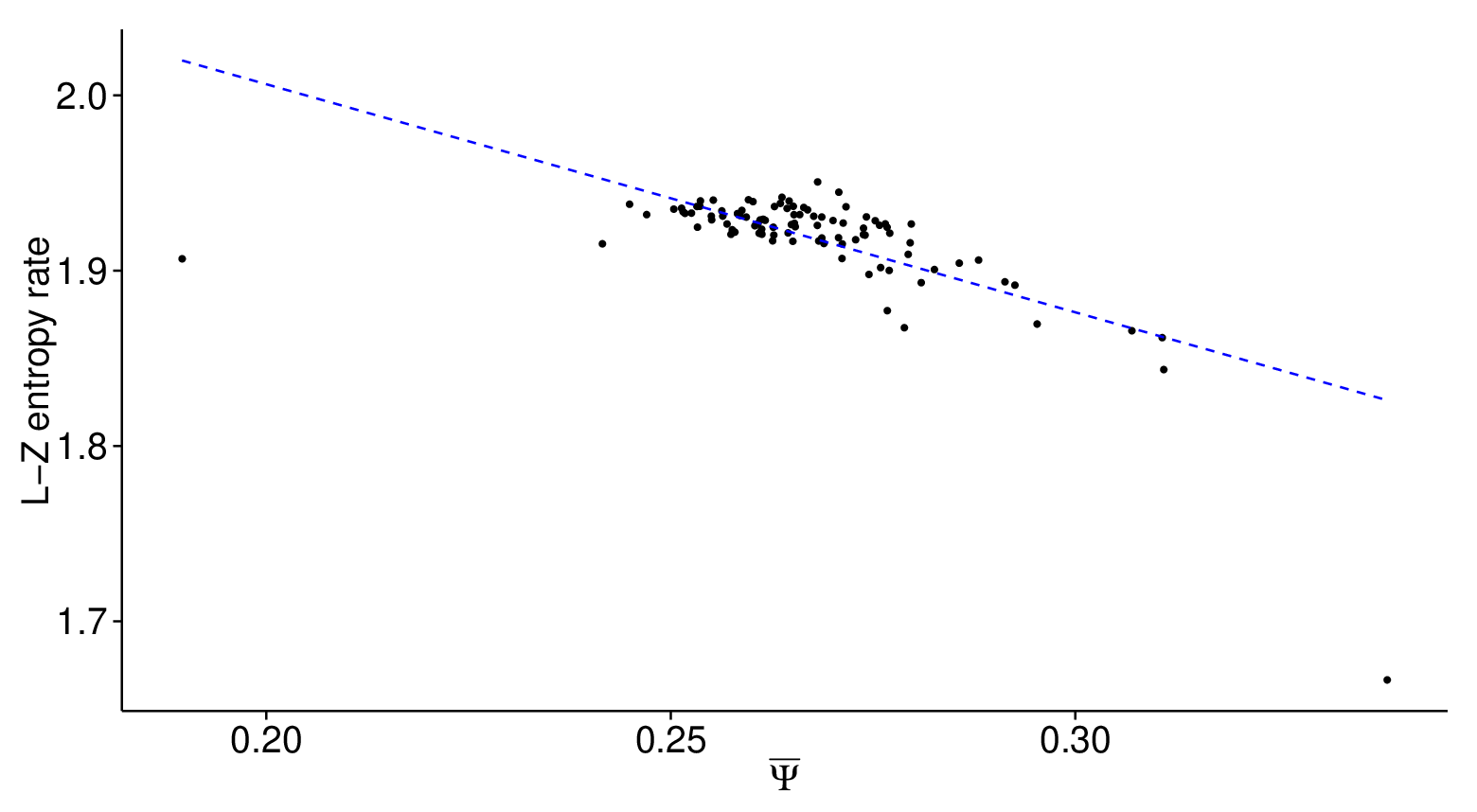}}%
\subfloat[][]{%
\label{fig:ND4-d}%
\includegraphics[width=0.5\textwidth]{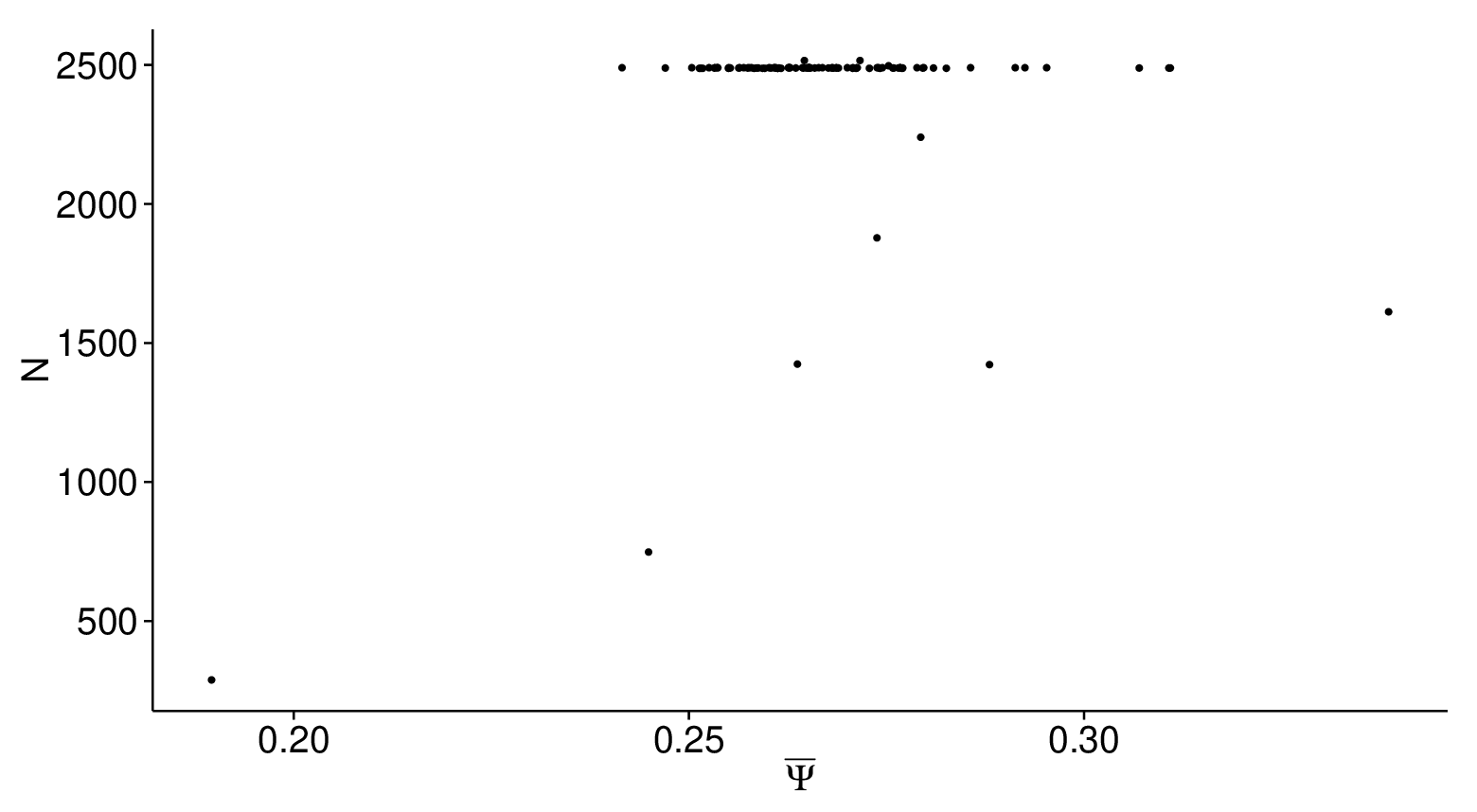}}%
\caption[ND4d]{Maximum Entropy Production Principle (MEPP) testing results for New York Stock Exchange daily logarithmic returns (discretised into four quartiles) database of 98 stocks:
\subref{fig:ND4-a} Estimate of the percentage of times the MEPP correctly predicts the next price move ($\Psi$) averaged over all studied stocks vs the window length parameter ($\mu$), together with error bars of one standard deviation, and two lines denoting values for random guessing the next price move (below unadjusted, above adjusted);
\subref{fig:ND4-b} Histogram of $\Psi$ averaged over all values of $\mu$ together with vertical lines for unadjusted (left) and adjusted (right) values for random guessing;
\subref{fig:ND4-c} Scatterplot of $\Psi$ averaged over all values of $\mu$ vs Lempel--Ziv entropy rate for the whole time series, with linear regression fitted;
\subref{fig:ND4-d} Scatterplot of $\Psi$ averaged over all values of $\mu$ vs the length of the whole time series. Both the average and a great majority of individual stocks are above the values for random chance, supporting our hypothesis. Additionally, the values of $\Psi$ are strongly positively correlated with the predictability of the price formation processes, but the results do not depend on the size of the studied time series.}%
\label{fig:ND4}%
\end{figure*}

\begin{figure*}%
\centering
\subfloat[][]{%
\label{fig:NI4-a}%
\includegraphics[width=0.5\textwidth]{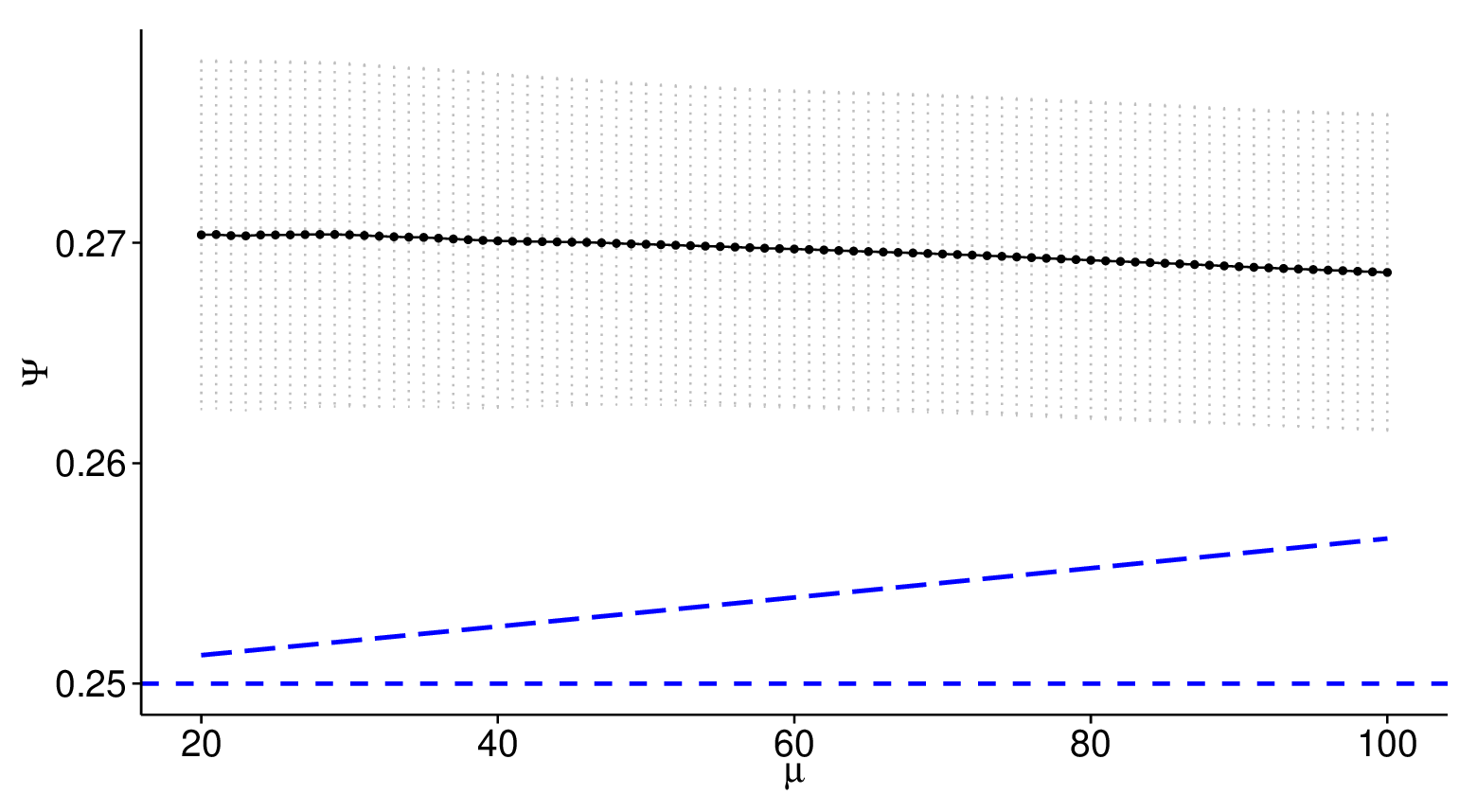}}%
\subfloat[][]{%
\label{fig:NI4-b}%
\includegraphics[width=0.5\textwidth]{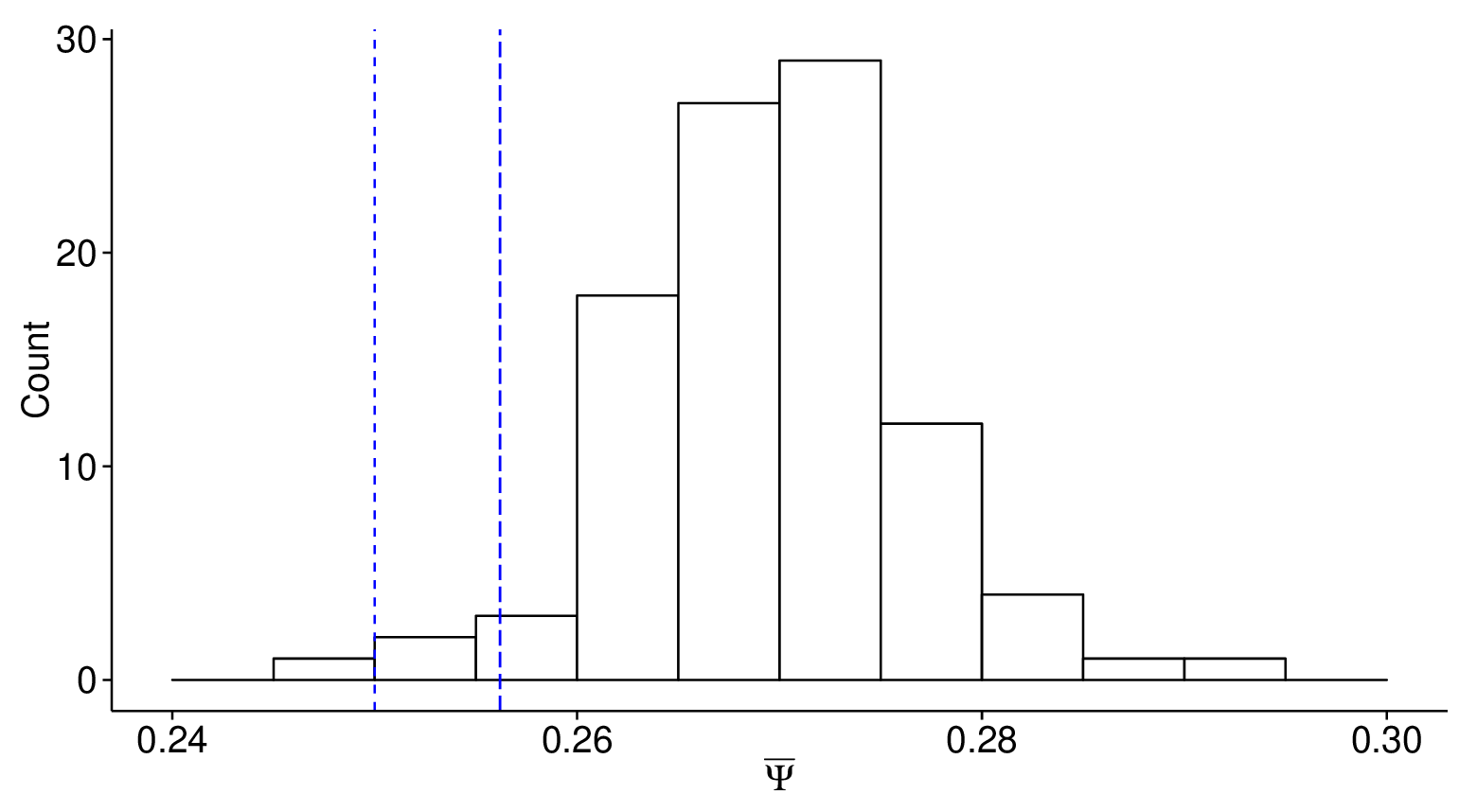}}\\
\subfloat[][]{%
\label{fig:NI4-c}%
\includegraphics[width=0.5\textwidth]{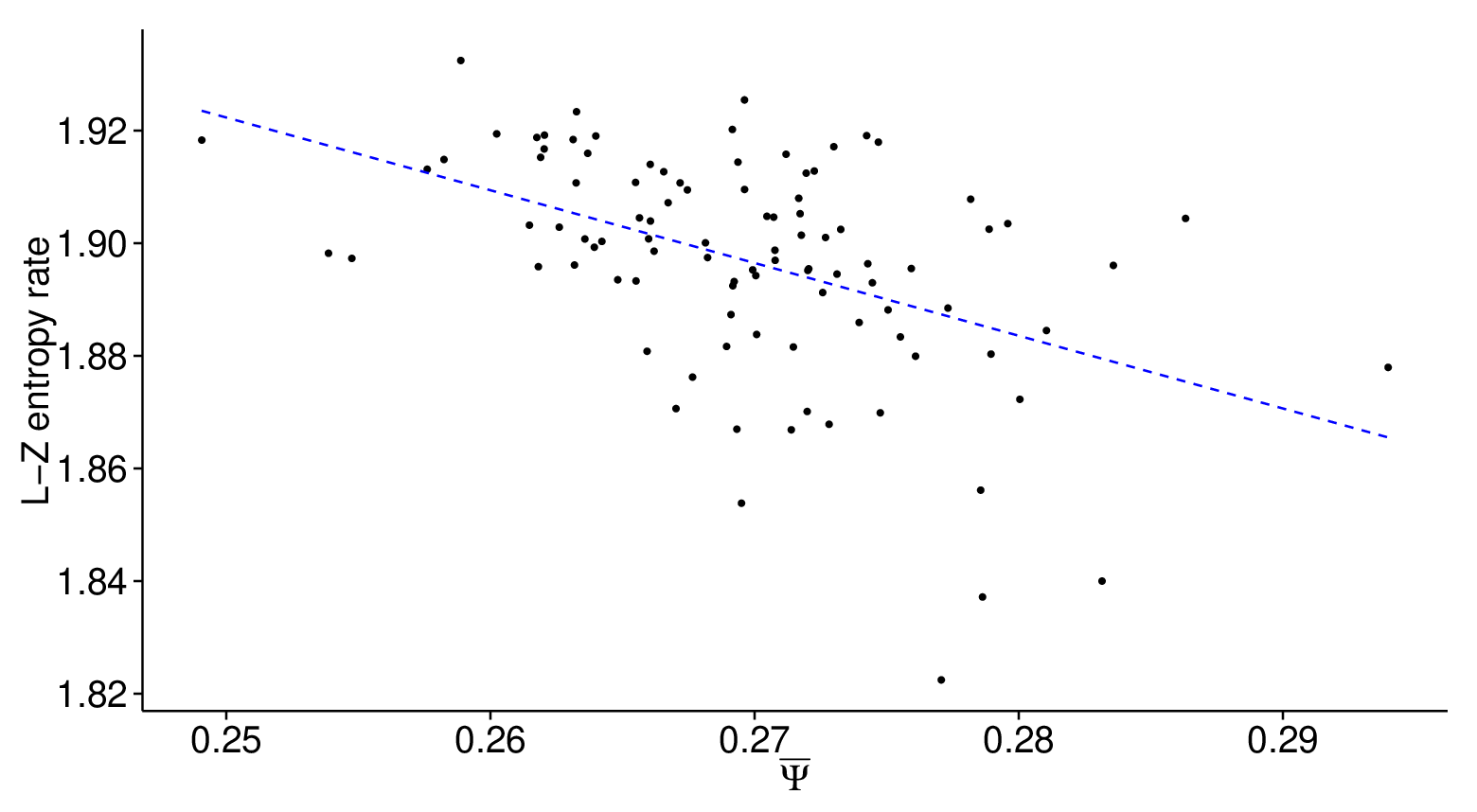}}%
\subfloat[][]{%
\label{fig:NI4-d}%
\includegraphics[width=0.5\textwidth]{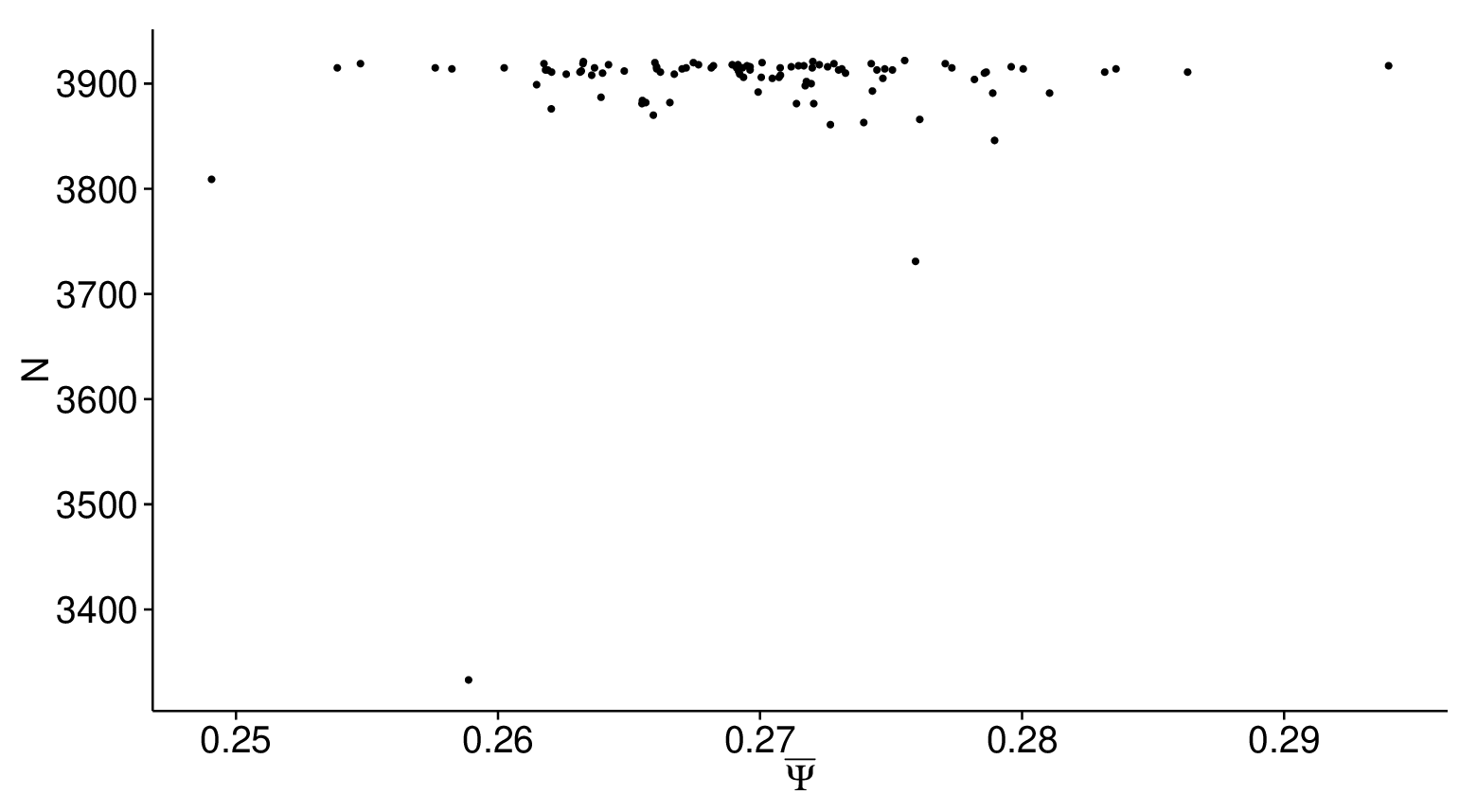}}%
\caption[NI4d]{Maximum Entropy Production Principle (MEPP) testing results for New York Stock Exchange intraday logarithmic returns (discretised into four quartiles) database of 98 stocks:
\subref{fig:NI4-a} Estimate of the percentage of times the MEPP correctly predicts the next price move ($\Psi$) averaged over all studied stocks vs the window length parameter ($\mu$), together with error bars of one standard deviation, and two lines denoting values for random guessing the next price move (below unadjusted, above adjusted);
\subref{fig:NI4-b} Histogram of $\Psi$ averaged over all values of $\mu$ together with vertical lines for unadjusted (left) and adjusted (right) values for random guessing;
\subref{fig:NI4-c} Scatterplot of $\Psi$ averaged over all values of $\mu$ vs Lempel--Ziv entropy rate for the whole time series, with linear regression fitted;
\subref{fig:NI4-d} Scatterplot of $\Psi$ averaged over all values of $\mu$ vs the length of the whole time series. Both the average and a great majority of individual stocks are above the values for random chance, supporting our hypothesis. Additionally, the values of $\Psi$ are strongly positively correlated with the predictability of the price formation processes, but the results do not depend on the size of the studied time series.}%
\label{fig:NI4}%
\end{figure*}

In figures~\ref{fig:WI4}, \ref{fig:ND4}, and \ref{fig:NI4} we show the same results for Warsaw Stock Exchange intraday logarithmic returns (discretised into four quartiles) database of 707 stocks, New York Stock Exchange daily logarithmic returns (discretised into four quartiles) database of 98 stocks, and New York Stock Exchange intraday logarithmic returns (discretised into four quartiles) database of 98 stocks respectively.

\begin{figure*}%
\centering
\subfloat[][]{%
\label{fig:dist4-a}%
\includegraphics[width=0.5\textwidth]{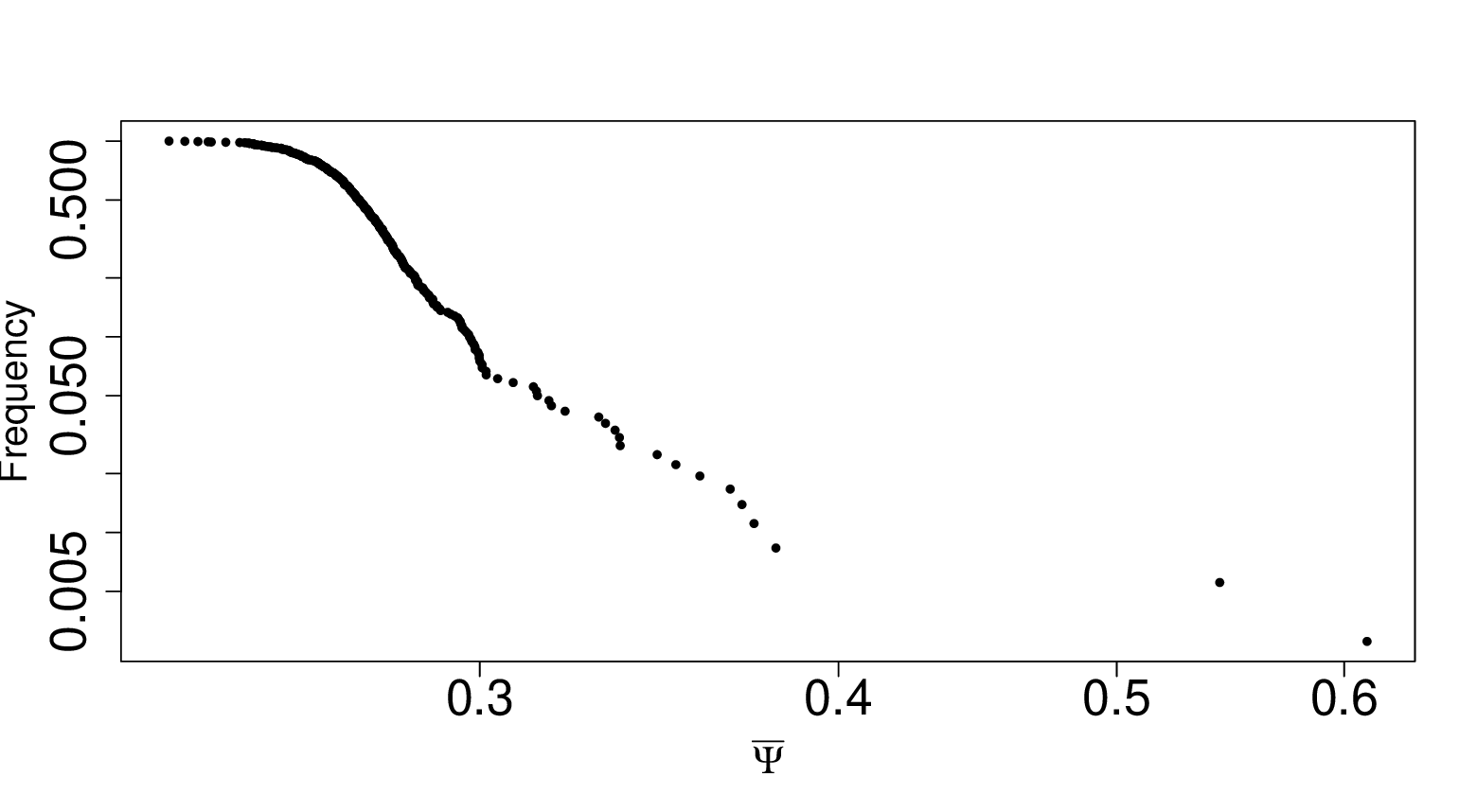}}%
\subfloat[][]{%
\label{fig:dist4-b}%
\includegraphics[width=0.5\textwidth]{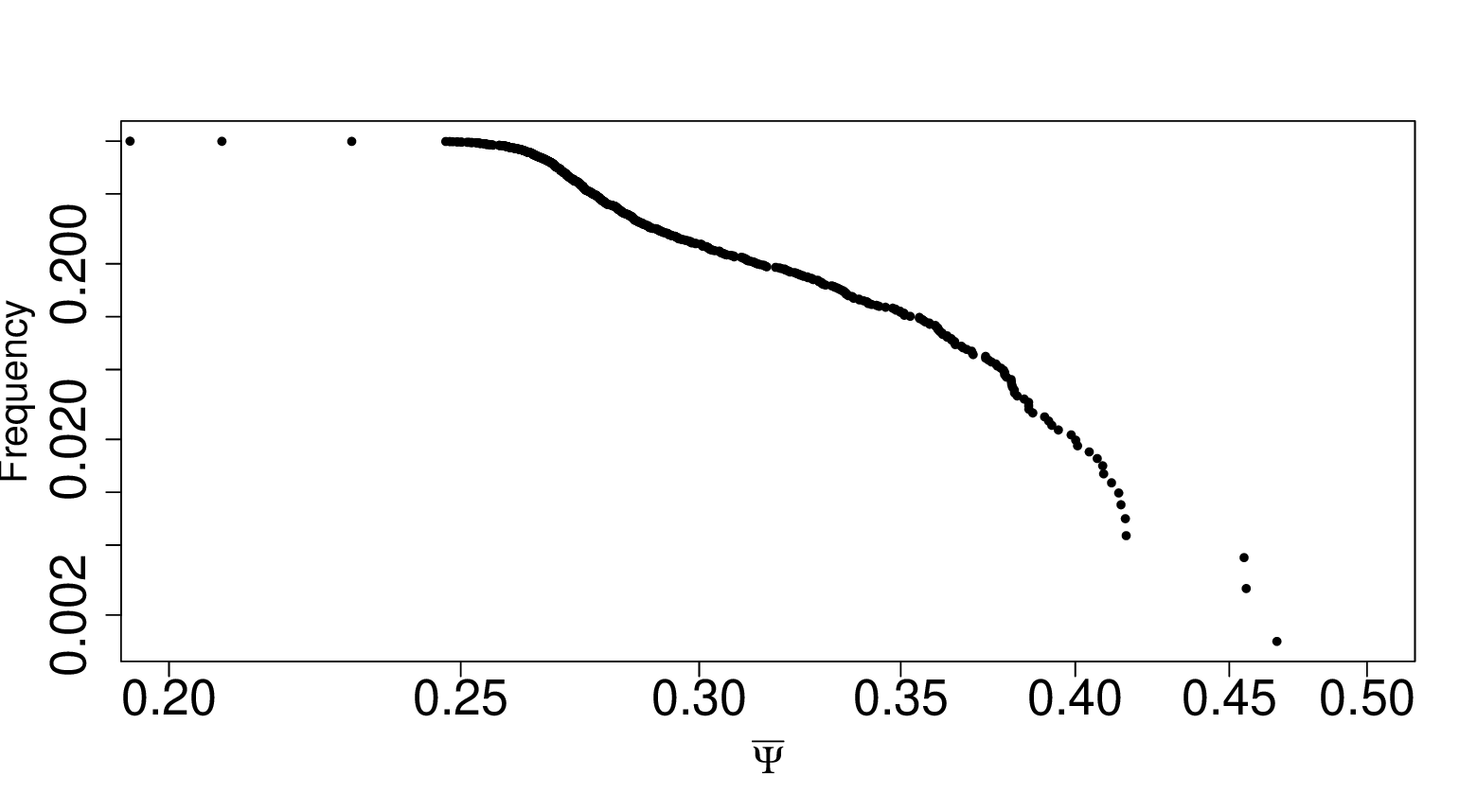}}\\
\subfloat[][]{%
\label{fig:dist4-c}%
\includegraphics[width=0.5\textwidth]{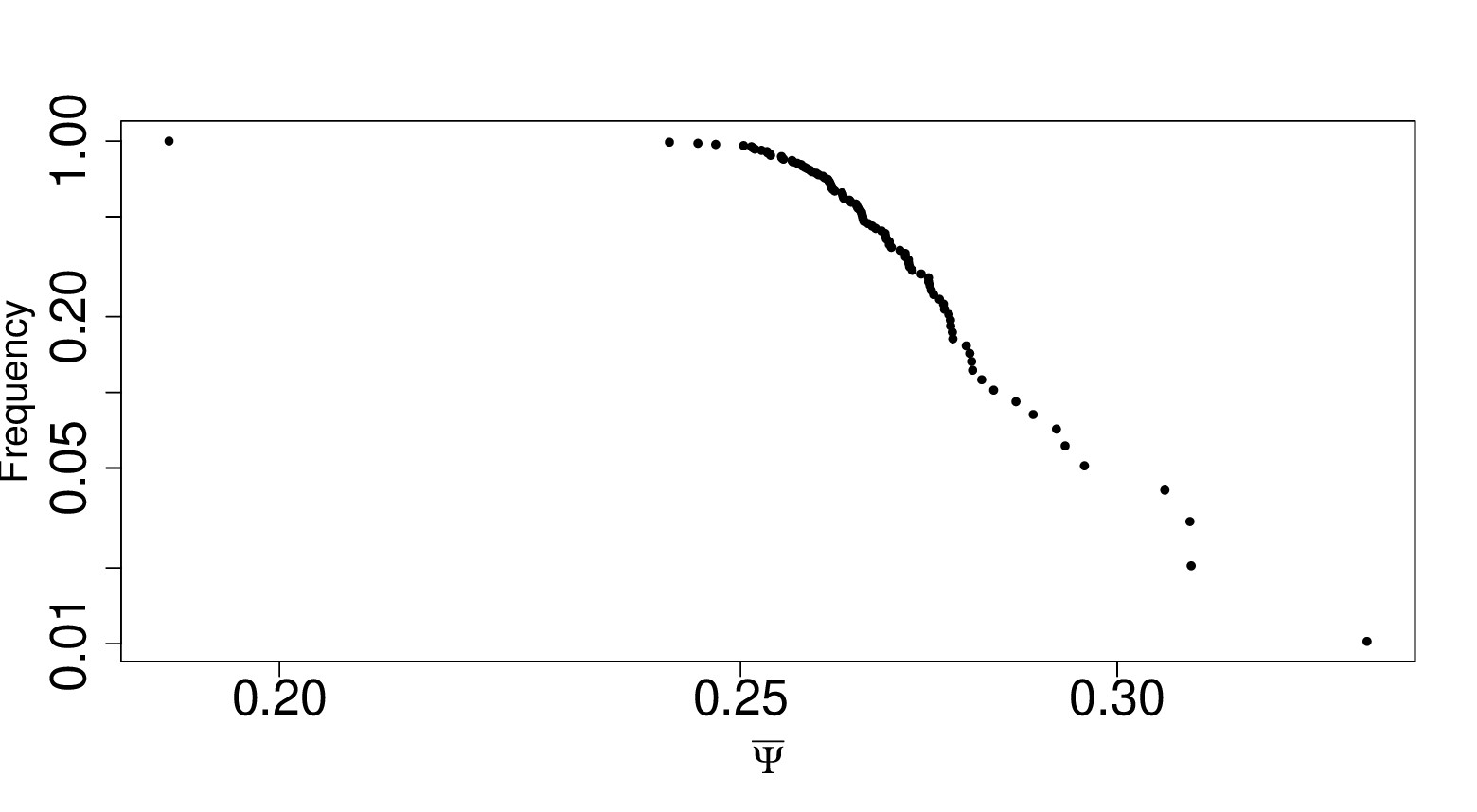}}%
\subfloat[][]{%
\label{fig:dist4-d}%
\includegraphics[width=0.5\textwidth]{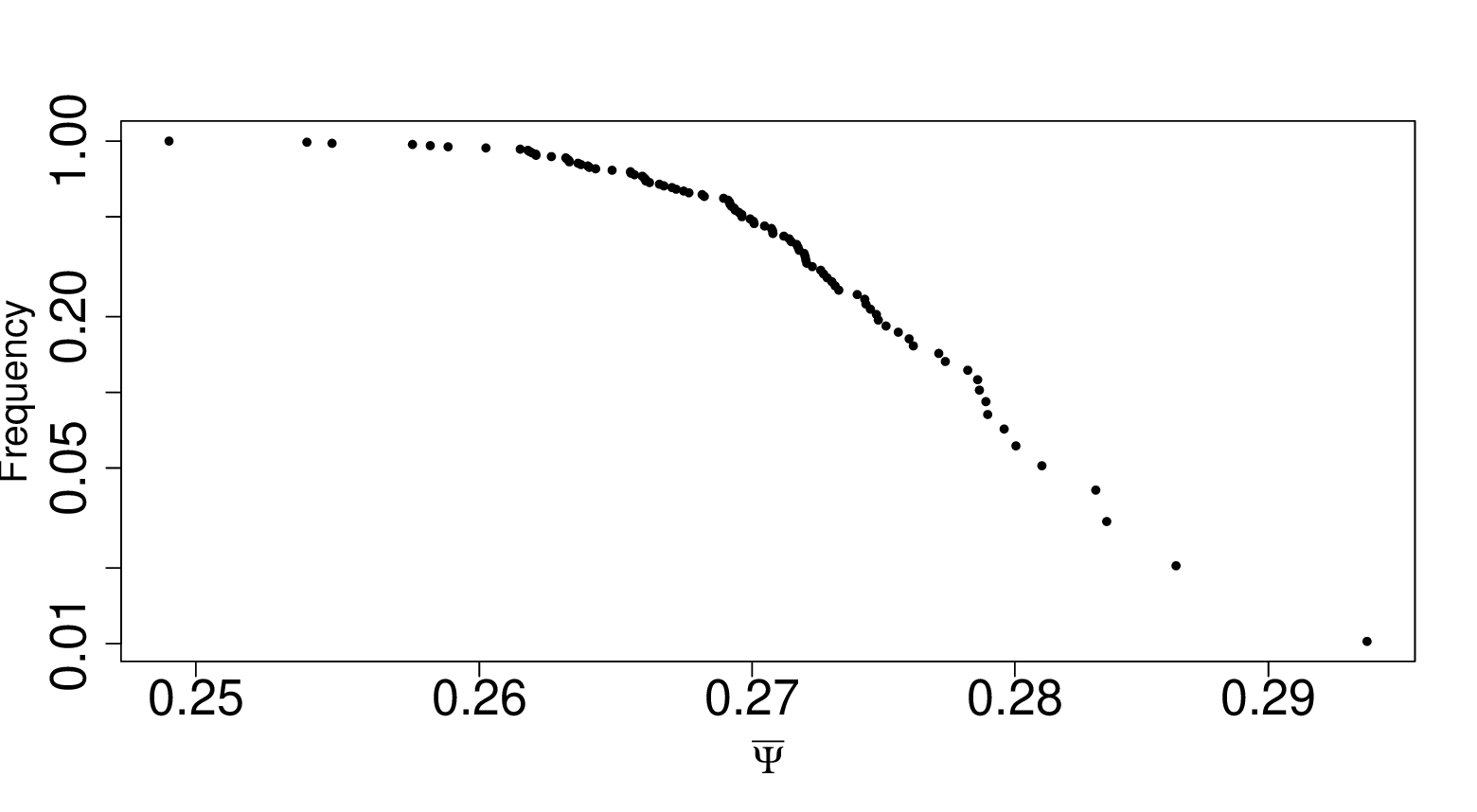}}%
\caption[Degree]{The distributions of the estimates of the percentage of times the Maximum Entropy Production Principle (MEPP) correctly predicts the next price move ($\Psi$), averaged over all values of the window length $\mu$, for logarithmic returns (discretised into four quartiles) for:
\subref{fig:dist4-a} Warsaw Stock Exchange daily database of 360 stocks;
\subref{fig:dist4-b} Warsaw Stock Exchange intraday database of 707 stocks;
\subref{fig:dist4-c} New York Stock Exchange daily database of 98 stocks;
\subref{fig:dist4-d} and New York Stock Exchange intraday database of 98 stocks. The distributions show signs of fat tails, particularly seen for the high values of $\Psi$. It is clear that almost the whole mass of the distributions is above $0.25$, or the estimate for random guessing the next price change, thus supporting our hypothesis of the MEPP within financial markets.}%
\label{fig:dist4}%
\end{figure*}

In figure~\ref{fig:dist4} we show the distributions of the estimates of the percentage of times the Maximum Entropy Production Principle (MEPP) correctly predicts the next price move ($\Psi$) averaged over all values of the parameter $\mu$ for logarithmic returns (discretised into four quartiles) for Warsaw Stock Exchange daily logarithmic returns (discretised into four quartiles) database of 360 stocks (figure~\ref{fig:dist4-a}), Warsaw Stock Exchange intraday logarithmic returns (discretised into four quartiles) database of 707 stocks (figure~\ref{fig:dist4-b}), New York Stock Exchange daily logarithmic returns (discretised into four quartiles) database of 98 stocks (figure~\ref{fig:dist4-c}), and New York Stock Exchange intraday logarithmic returns (discretised into four quartiles) database of 98 stocks (figure~\ref{fig:dist4-d}). These allow us too see where the mass of the distribution is, and whether the distributions have fat tails, which could be helpful for analysts and traders.

\section{Discussion}

First and foremost, we want to confirm or deny the hypothesis of the Maximum Entropy Production Principle for stock returns. Results presented in figures~\ref{fig:WD4}--\ref{fig:NI4} support the hypothesis as outlined in equations~\eqref{eq:hyp} and \eqref{eq:hyp2}. To decide whether we need the adjustment mentioned above, we observe how the results ($\Psi$) change with various values of window length $\mu$ and time series size $N$. We find that $\Psi$ does not depend on either of these, with a small exception for low values of $\mu$ where statistical noise changes the results (although only slightly). If the scenario of described over-representation of the guessed state in the studied time series were affecting our analysis we would expect to see higher values of $\Psi$ for higher values of $\mu$ and lower values of N, which is not the case. Therefore we conclude that it is safe to use the unadjusted hypothesis as defined in equation~\eqref{eq:hyp}. At the same time we note that using the hypothesis as defined in equation~\eqref{eq:hyp2} would leave the conclusions essentially the same. Having chosen the hypothesis to test against, we look at the average value of $\Psi$ we have obtained in the four studied sets, and compare it with the values we would expect for random choice ($1/\Omega=0.25$). For Warsaw Stock Exchange we have obtained average $\Psi$ of $0.277$ and $0.291$ for daily and intraday logarithmic returns respectively. These are averaged over all studied stocks, and also all studied values of the window length $\mu$. As mentioned above, an average over all values of $\mu$ is informative as there is practically no dispersion in these results, with standard deviations of $0.0011$ and $0.0014$ respectively. For New York Stock Exchange we have obtained average $\Psi$ of $0.267$ and $0.270$ for daily and intraday logarithmic returns respectively. In these there also is practically no dispersion, with standard deviations of $0.0025$ and $0.0054$ respectively. Additionally, for Warsaw Stock Exchange only an average of 14 out of 360 stocks show average $\Psi$ below the $0.25$ threshold (between 10 and 21 for various values of $\mu$) for the daily log returns, and only an average of 10 out of 707 stocks (between 7 and 11) for the intraday log returns. For New York Stock Exchange only an average of 6 out of 98 stocks show average $\Psi$ below the $0.25$ threshold (between 4 and 12 for various values of $\mu$) for the daily log returns, and only an average of 1 out of 98 stocks (between 0 and 1) for the intraday log returns. From these results we get a clear support for the hypothesis of the application of the MEPP to stock returns as defined in equation~\eqref{eq:hyp}. We get only a handful of stocks for which the estimated $\Psi$ is under the threshold. These results are particularly strong for intraday logarithmic returns, which is not surprising as these have been known to be more predictable.

Second, we take a look at whether the Maximum Entropy Production Principle allows us to guess future price changes more accurately for these price formation processes which are more predictable in the sense of information theory (entropy rate for the whole studied series). In previous studies \citep{Navet:2008,Fiedor:2014}, standard ways of predicting future price changes have been shown not to use the predictability (structural complexity) of the time series describing logarithmic returns. We would like to find an approach which is capable of using this information. And, as can be seen in the figures described above, there is a strong negative correlation between $\Psi$ and entropy rate for the studied time series describing logarithmic returns. In other words, there is a strong positive correlation between the explanatory power of the Maximum Entropy Production Principle and the predictability of the underlying price formation processes. The Pearson's correlation coefficient between the $\Psi$ averaged over all studied values of $\mu$ and the Lempel--Ziv entropy rate for all studied stocks are equal to $-0.937$ and $-0.628$ for Warsaw Stock Exchange daily and intraday data, together with $-0.678$ and $-0.465$ for New York Stock Exchange daily and intraday data respectively. Thus we are able to conclude that the use of the MEPP as a tool for predicting future price changes of stocks uses the predictability of the price formation processes, contrary to popular methods used in trading algorithms, such as the mean reversal strategy.

Third, we take a look at the distributions of the estimates of the percentage of times the Maximum Entropy Production Principle correctly predicts the next price move ($\Psi$), averaged over all values of the window length $\mu$, for logarithmic returns within the four studied sets. As can be seen in figure~\ref{fig:dist4}, almost the whole mass of the distribution is above $0.25$ for all four studied sets of stocks, thus supporting our hypothesis of the Maximum Entropy Production Principle within financial markets. Additionally, we observe that for the large values of $\Psi$ these distributions are characterised by fat tails, roughly approximating log-normal distribution. This is interesting and important, as the analysts and traders may concentrate on the handful of stocks with the largest predictability as defined by their Lempel--Ziv entropy rate, which are then likely to give the best results when applying the Maximum Entropy Production Principle for prediction purposes.

Finally, we also note that this method, similarly to earlier studies of structural complexity of stock returns, uses volatility levels to distinguish patterns. So while this method should give good results if $\Omega$ is larger than four, it will not give very good results when we apply it to stock returns discretised with $\Omega=2$, that is only distinguishing positive and negative logarithmic returns. Within such framework there is not enough patterns within the data (which are mostly based on volatility clustering) for the predictability to be useful for analysis or predictions based on information theory.

\section{Conclusions}

In this paper we have introduced the concept of the Maximum Entropy Production Principle as applied to financial markets, and logarithmic stock returns in particular. We propose that such a principle governs the financial markets, but that it is bounded by a large number of unknown constraints, so that in practice it can only be observed as a statistical tendency, and not an immutable law. Thus our testing it, on the basis of databases containing historical daily and intraday logarithmic returns for stock exchanges in Warsaw and New York, can only partially confirm the hypothesis of such principle. As is often the case with complex systems, a full proof of such a simple principle will be very hard to obtain. Nonetheless, the results confirmed our hypothesis and we have found that the application of such principle to the prediction of future stock returns gives consistently better returns than would be obtained by random chance. Additionally, the performance of such prediction, based on the Maximum Entropy Production Principle, is strongly dependent on the structural complexity, or model-free predictability, of the underlying price formation processes. This is important, as traditional trading algorithms do not possess such a characteristic. Together with the fact that the distribution of the performance of the Maximum Entropy Production Principle has fat tails, this allows analysts to pick stocks for which the underlying price formation processes are the most predictable, and concentrate their analysis or trading on these, knowing that it is more likely these will produce better results with the use of the analysed principle. In particular this means using intraday stock returns, as these are more predictable than their daily counterparts. Future studies should look into the application of the Maximum Entropy Production Principle to other financial markets and instruments, such as currency exchange rates. Further studies should also try to develop the methodology in various ways, for example introducing the usage of permutation entropy for the purposes of such an analysis, or apply the same principle to time series describing logarithmic returns with normalised volatility. Studies should also be performed to establish whether it is possible to find the constraints to this principle on the markets, as mentioned above.

\vspace{-1mm}

\vspace{-1mm}

\bibliographystyle{rQUF}
\bibliography{prace}

\begin{thebibliography}{39}
\providecommand{\natexlab}[1]{#1}
\providecommand{\noopsort}[1]{}
\providecommand{\printfirst}[2]{#1}
\providecommand{\singleletter}[1]{#1}
\providecommand{\switchargs}[2]{#2#1}

\bibitem[\protect\citeauthoryear{Beretta}{2010}]{Beretta:2010}
Beretta, G., {Maximum entropy production rate in quantum thermodynamics}.
  {\itshape Journal of Physics: Conference Series}, 2010, \textbf{237}, 012004.

\bibitem[\protect\citeauthoryear{Boltzmann}{1866}]{Boltzmann:1866}
Boltzmann, L., {Uber die Mechanische Bedeutung des Zweiten Hauptsatzes der
  Warmetheorie}. {\itshape {Wiener Berichte}}, 1866, \textbf{53}, 195--220.

\bibitem[\protect\citeauthoryear{Clausius}{1854}]{Clausius:1854}
Clausius, R., {Uber eine veranderte Form des zweiten Hauptsatzes der
  mechanischen Wärmetheoriein}. {\itshape {Annalen der Physik und Chemie}},
  1854, \textbf{93}, 481--506.

\bibitem[\protect\citeauthoryear{Cover and Thomas}{1991}]{Cover:1991}
Cover, T. and Thomas, J., {\itshape Elements of Information Theory}, 1991
  (John Wiley \& Sons: New York).

\bibitem[\protect\citeauthoryear{Doganaksoy and
  Gologlu}{2006}]{Doganaksoy:2006}
Doganaksoy, A. and Gologlu, F., {On Lempel-Ziv Complexity of Sequences}.
  {\itshape {Lecture Notes in Computer Science}}, 2006, \textbf{4086},
  180--189.

\bibitem[\protect\citeauthoryear{Dugdale}{1996}]{Dugdale:1996}
Dugdale, J., {\itshape {Entropy and its Physical Meaning}}, 1996  ({Taylor and
  Francis}: London).

\bibitem[\protect\citeauthoryear{Farah {\itshape{et~al.}}}{1995}]{Farah:1995}
Farah, M., Noordewier, M., Savari, S., Shepp, L., Wyner, A. and Ziv, J., {On
  the entropy of DNA: algorithms and measurements based on memory and rapid
  convergence}. In {\itshape Proceedings of the }{\itshape SODA'95: Proceedings
  of the Sixth Annual ACM-SIAM Symposium on Discrete Algorithms}, pp. 48--57,
  1995  (San Francisco).

\bibitem[\protect\citeauthoryear{Fiedor}{2014}]{Fiedor:2014}
Fiedor, P., {Frequency Effects on Predictability of Stock Returns}. In
  {\itshape Proceedings of the }{\itshape {Proceedings of the IEEE
  Computational Intelligence for Financial Engineering \& Economics 2014}},
  edited by A.~Serguieva, D.~Maringer, V.~Palade and R.J. Almeida, pp.
  247--254, 2014  ({IEEE}: London).

\bibitem[\protect\citeauthoryear{Gao {\itshape{et~al.}}}{2006}]{Gao:2006}
Gao, Y., Kontoyiannis, I. and Bienenstock, E., From the entropy to the
  statistical structure of spike trains. In {\itshape Proceedings of the
  }{\itshape IEEE International Symposium on Information Theory}, pp. 645--649,
  2006  (Seattle).

\bibitem[\protect\citeauthoryear{Gao {\itshape{et~al.}}}{2008}]{Gao:2008}
Gao, Y., Kontoyiannis, I. and Bienenstock, E., Estimating the Entropy of Binary
  Time Series: Methodology, Some Theory and a Simulation Study. {\itshape
  Entropy}, 2008, \textbf{10}, 71--99.

\bibitem[\protect\citeauthoryear{Gheorghiu-Svirschevski}{2001}]{Gheorghiu:2001}
Gheorghiu-Svirschevski, S., {Nonlinear quantum evolution with maximal entropy
  production}. {\itshape Physical Review A}, 2001, \textbf{63}, 022105.

\bibitem[\protect\citeauthoryear{Gibbs}{1902}]{Gibbs:1902}
Gibbs, J., {\itshape {Elementary Principles in Statistical Mechanics --
  Developed with Especial Reference to the Rational Foundation of
  Thermodynamics}}, 1902  ({C. Scribner's Sons}: New York).

\bibitem[\protect\citeauthoryear{Gunzig
  {\itshape{et~al.}}}{1987}]{Prigogine:1987}
Gunzig, E., Geheniau, J. and Prigogine, I., {Entropy and cosmology}. {\itshape
  Nature}, 1987, \textbf{330}, 621--624.

\bibitem[\protect\citeauthoryear{Henin and Prigogine}{1974}]{Henin:1974}
Henin, F. and Prigogine, I., Entropy, dynamics, and molecular chaos. {\itshape
  Proceedings of the National Academy of Sciences}, 1974, \textbf{71},
  2618--2622.

\bibitem[\protect\citeauthoryear{Ingber}{1984}]{Ingber:1984}
Ingber, L., {Statistical mechanics of nonlinear nonequilibrium financial
  markets}. {\itshape {Mathematical Modelling}}, 1984, \textbf{5}, 343--361.

\bibitem[\protect\citeauthoryear{Keizer}{1987}]{Keizer:1987}
Keizer, J., {\itshape {Statistical Thermodynamics of Nonequilibrium
  Processes}}, 1987  ({Springer-Verlag}: {New York}).

\bibitem[\protect\citeauthoryear{Kennel {\itshape{et~al.}}}{2005}]{Kennel:2005}
Kennel, M., Shlens, J., Abarbanel, H. and Chichilnisky, E., {Estimating entropy
  rates with Bayesian confidence intervals}. {\itshape {Neural Computation}},
  2005, \textbf{17}, 1531--1576.

\bibitem[\protect\citeauthoryear{Kolmogorov}{1959}]{Kolmogorov:1959}
Kolmogorov, A.N., Entropy per unit time as a metric invariant of automorphism.
  {\itshape Doklady Akademii Nauk SSSR}, 1959, \textbf{124}, 754--755.

\bibitem[\protect\citeauthoryear{Kontoyiannis}{1998}]{Kontoyiannis:1998}
Kontoyiannis, I., {Asymptotically Optimal Lossy Lempel-Ziv Coding}. In
  {\itshape Proceedings of the }{\itshape IEEE International Symposium on
  Information Theory}, 1998  (Cambridge).

\bibitem[\protect\citeauthoryear{Kontoyiannis
  {\itshape{et~al.}}}{1998}]{Kontoyiannis:1998a}
Kontoyiannis, I., Algoet, P., Suhov, Y. and Wyner, A., {Nonparametric entropy
  estimation for stationary processes and random fields, with applications to
  English text}. {\itshape {IEEE Transactions on Information Theory}}, 1998,
  \textbf{44}, 1319--1327.

\bibitem[\protect\citeauthoryear{Lempel and Ziv}{1977}]{Lempel:1977}
Lempel, A. and Ziv, J., {A Universal Algorithm for Sequential Data
  Compression}. {\itshape {IEEE Transactions on Information Theory}}, 1977,
  \textbf{23}, 337--343.

\bibitem[\protect\citeauthoryear{Leonardi}{2010}]{Leonardi:2010}
Leonardi, F., Some upper bounds for the rate of convergence of penalized
  likelihood context tree estimators. {\itshape Brazilian Journal of
  Probability and Statistics}, 2010, \textbf{24}, 321--336.

\bibitem[\protect\citeauthoryear{Louchard and
  Szpankowski}{1997}]{Louchard:1997}
Louchard, G. and Szpankowski, W., {On the average redundancy rate of the
  Lempel-Ziv code}. {\itshape {IEEE Transactions on Information Theory}}, 1997,
  \textbf{43}, 2--8.

\bibitem[\protect\citeauthoryear{Martyushev and
  Seleznev}{2006}]{Martyushev:2006}
Martyushev, L. and Seleznev, V., {Maximum entropy production principle in
  physics, chemistry and biology}. {\itshape Physical Reports}, 2006,
  \textbf{426}, 1--45.

\bibitem[\protect\citeauthoryear{Martyushev
  {\itshape{et~al.}}}{2000}]{Martyushev:2000}
Martyushev, L., Seleznev, V. and Kuznetsova, I., {Application of the principle
  of maximum entropy production to the analysis of the morphological stability
  of a growing crystal}. {\itshape Journal of Experimental and Theoretical
  Physics}, 2000, \textbf{91}, 132--143.

\bibitem[\protect\citeauthoryear{Navet and Chen}{2008}]{Navet:2008}
Navet, N. and Chen, S.H., {On Predictability and Profitability: Would GP
  Induced Trading Rules be Sensitive to the Observed Entropy of Time Series?}.
  In {\itshape {Natural Computing in Computational Finance}}, edited by
  T.~Brabazon and M.~O'Neill, Vol.  100 of {\itshape Studies in Computational
  Intelligence}, 2008  (Springer: Berlin Heidelberg).

\bibitem[\protect\citeauthoryear{Officer}{1972}]{Officer:1972}
Officer, R., {The Distribution of Stock Returns}. {\itshape {Journal of the
  American Statistical Association}}, 1972, \textbf{340}, 807--812.

\bibitem[\protect\citeauthoryear{Onsager}{1931}]{Onsager:1931}
Onsager, L., {Reciprocal relations in irreversible processes}. {\itshape
  {Physical Review}}, 1931, \textbf{37}, 405--426.

\bibitem[\protect\citeauthoryear{Ozawa {\itshape{et~al.}}}{2003}]{Ozawa:2003}
Ozawa, H., Ohmura, A., Lorenz, R. and Pujol, T., {The second law of
  thermodynamics and the global climate system: A review of the maximum entropy
  production principle}. {\itshape Review of Geophysics}, 2003, \textbf{41},
  1018--1042.

\bibitem[\protect\citeauthoryear{Prigogine}{1950}]{Prigogine:1950}
Prigogine, I., Sur les fluctuations de l'equilibre chimique. {\itshape
  Physica}, 1950, \textbf{16}, 134--136.

\bibitem[\protect\citeauthoryear{Prigogine}{1947}]{Prigogine:1947}
Prigogine, I., {Etude thermodynamique des phenomenes irreversibles}. PhD
  thesis, Universite libre de Bruxelles, 1947.

\bibitem[\protect\citeauthoryear{Shannon}{1948}]{Shannon:1948}
Shannon, C.E., {A Mathematical Theory of Communication}. {\itshape Bell Systems
  Techical Journal}, 1948, \textbf{27}, 379--423, 623--656.

\bibitem[\protect\citeauthoryear{Sinai}{1959}]{Sinai:1959}
Sinai, Y., {On the Notion of Entropy of a Dynamical System}. {\itshape {Doklady
  Akademii Nauk SSSR}}, 1959, \textbf{124}, 768--771.

\bibitem[\protect\citeauthoryear{Starica and Granger}{2005}]{Starica:2005}
Starica, C. and Granger, C., {Nonstationarities in stock returns}. {\itshape
  {The Review of Economics and Statistics}}, 2005, \textbf{87}, 503--522.

\bibitem[\protect\citeauthoryear{Steuer {\itshape{et~al.}}}{2001}]{Steuer:2001}
Steuer, R., Molgedey, L., Ebeling, W. and Jiménez-Monta\~{n}o, M., {Entropy and
  Optimal Partition for Data Analysis}. {\itshape European Physical Journal B},
  2001, \textbf{19}, 265--269.

\bibitem[\protect\citeauthoryear{Voit}{2001}]{Voit:2001}
Voit, J., {\itshape {The Statistical Mechanics of Financial Markets}}, 2001
  ({Springer}: London).

\bibitem[\protect\citeauthoryear{Willems}{1998}]{Willems:1998}
Willems, F., The context-tree weighting method: Extensions. {\itshape {IEEE
  Transactions on Information Theory}}, 1998, \textbf{44}, 792--798.

\bibitem[\protect\citeauthoryear{Willems
  {\itshape{et~al.}}}{1996}]{Willems:1996}
Willems, F., Shtarkov, Y. and Tjalkens, T., Context weighting for general
  finite-context sources. {\itshape {IEEE Transactions on Information Theory}},
  1996, \textbf{42}, 1514--1520.

\bibitem[\protect\citeauthoryear{Willems
  {\itshape{et~al.}}}{1995}]{Willems:1995}
Willems, F., Shtarkov, Y. and Tjalkens, T., {The Context-Tree Weighting Method:
  Basic Properties}. {\itshape {IEEE Transactions on Information Theory}},
  1995, \textbf{41}, 653--664.

\end{thebibliography}
\end{document}